\documentclass[%
 aip,
 amsmath,amssymb,
preprint,%
]{revtex4-1}

\usepackage{graphicx}
\usepackage{dcolumn}
\usepackage{bm}

\usepackage[utf8]{inputenc}
\usepackage[T1]{fontenc}
\usepackage{mathptmx}

\usepackage{hyperref}
\begin{document}

\preprint{AIP/123-QED}

\title{The Snowball Earth transition in a climate model with drifting parameters}

\author{Bálint Kaszás}
\affiliation{
Institute for Theoretical Physics, Eötvös Loránd University, Pázmány Péter Sétány 1/A, H-1117 Budapest,
Hungary}%
\author{Tímea Haszpra}%
\affiliation{
Institute for Theoretical Physics, Eötvös Loránd University, Pázmány Péter Sétány 1/A, H-1117 Budapest,
Hungary}%
\affiliation{MTA-ELTE Theoretical Physics Research Group, Pázmány Péter Sétány 1/A, H-1117 Budapest, Hungary}
\author{Mátyás Herein}
\affiliation{
Institute for Theoretical Physics, Eötvös Loránd University, Pázmány Péter Sétány 1/A, H-1117 Budapest,
Hungary}%
\affiliation{MTA-ELTE Theoretical Physics Research Group, Pázmány Péter Sétány 1/A, H-1117 Budapest, Hungary}
\date{\today}

\begin{abstract}
Using an intermediate complexity climate model (Planet Simulator), we investigate the so-called Snowball Earth transition. For certain values of the solar constant, the climate system allows two different stable states: one of them is the Snowball Earth, covered by ice and snow, and the other one is today's climate.
In our setup, we consider the case when the climate system starts from its warm attractor (the stable climate we experience today), and the solar constant is decreased continuously in finite time, according to a parameter drift scenario, to a state, where only the Snowball Earth's attractor remains stable. This induces an inevitable transition, or climate tipping from the warm climate.
The reverse transition is also discussed. Increasing the solar constant back to its original value on individual simulations, we find that the system stays stuck in the Snowball state. However, using ensemble methods i.e., using an ensemble of climate realizations differing only slightly in their initial conditions we show that the transition from the Snowball Earth to the warm climate is also possible with a certain probability. From the point of view of dynamical systems theory, we can say that the system's snapshot attractor splits between the warm climate's and the Snowball Earth's attractor. 
\end{abstract}

\maketitle

\begin{quotation}
Ever since its discovery, the Snowball Earth, i.e. when the Earth's surface is nearly entirely frozen, received much attention within the climate science community. Much of the details of the transition to the planet's frozen state are still unexplored. Here, instead of focusing on the true Snowball events of Earth's history, we investigate the transition in an intermediate complexity climate model (PlaSim), with a continuously drifting solar constant (a hypothetical climate change scenario), in which a full return to the original value occurs. {\bf Using an ensemble based method we obtain both of the possible stable states as possible outcomes. We also show that the process is probabilistic and the probabilities of the corresponding outcomes are given by the ensemble's distribution. In addition, the third, unstable  state (referred to as the edge state) is also recovered.}\end{quotation}

\section{\label{sec:int}Introduction}
Snowball Earth refers to the planet's coldest possible global climate. In this state, the whole Earth, from the poles to the Equator, is covered in ice and snow. Since the thick ice covering the surface reflects much of the energy radiated by the Sun, the global average temperature is very low, around 220~K \cite{Harland1964}.

Modern findings suggest that during the Earth's history, there were periods, when such Snowball events occurred. For example, several traces of glacial activity point to the presence of glaciers along the so-called Paleoequator  \cite{Kirschvink1992}.

This  suggests that also the current configuration of the Earth system may be bistable, the two stable states being the Snowball state and our current climate. To better understand the phenomenon, there are simple models available that only take the global energy balance into account (Energy Balance Models). The first such attempts to understand snowball dynamics were made by Budyko \cite{Budyko1969} and Sellers \cite{Sellers1969}. These describe the Earth system's energy balance by a set of differential equations, that allow two stable equilibrium solutions. 

Qualitatively, the transition can be explained by the {\em ice-albedo feedback} \cite{Ghil1987}. Because of ice's higher albedo, if ice starts to accumulate on the planet, it causes even faster cooling. This is also true in reverse, i. e. melting ice decreases the overall albedo and causes a faster warming.

To start the feedback mechanism, (that is, to initiate a transition between the stable climates) global processes are necessary \cite{Pierrehumbert2011ClimateNeoproterozoic,  Feulner2015, tziperman2013}, which alter the amount of solar radiation absorbed by the surface.
For example, the eruption of a supervolcano or a series of volcanoes may set off the global freezing. It launches a huge amount of volcanic ash and sulfur into the atmosphere which ''shadows'' the planet, and causes the global temperature to decrease, see, e.g., \cite{Robock1995} for the Mount Pinatubo eruption in 1991.


The reverse process, the sudden melting of a Snowball Earth can be induced by greenhouse gases (e. g. CO$_2$). This can be also attributed to volcanic activity. Furthermore, the frozen planet prevents various processes that would extract CO$_2$ from the atmosphere, the most important of which are the outage of the entire biosphere and the frozen oceans \cite{Rose2017}.

On top of these natural causes, nowadays the idea of ''geoengineering'' \cite{geoengineer} emerged. Its main goal is to decelerate global warming by either reducing the concentration of greenhouse gases or by {\em artificially} reducing the surface's absorbed radiation. So far, it faces obvious technical difficulties. Nevertheless, possibly catastrophic consequences (which may result in a global cooling) cannot be ruled out.

Here, we follow a model-oriented approach. We investigate the transition between Earth's multiple equilibrium states using an intermediate complexity climate model. We induce the desired transition (both the planet's freezing, and then the melting) by changing one of the model's parameters in time. For simplicity, we choose the solar constant, which determines the amount of radiation that the Earth is subjected to. In the following, this parameter will be made {\em time dependent}, with a prescribed scenario.
It is important to emphasize that in reality, the solar constant is only a function of the Sun's activity and the distance from the Sun. In the last 400 years, its temporal variability was much smaller (about 0.2\% \cite{Steinhilber2012}) than what we will be considering in the main part of the text. Although there are certain processes in solar physics that may cause a considerable decrease in solar output, (see, e. g. the Faint Young Sun Paradox\cite{faintsun}), these happened on much longer time scales than those relevant for our purposes. 

However, in a model unable to describe volcanic activity, to compensate this, it is feasible to consider this parameter to change considerably. Especially, if the main driving force is considered to be the change in the radiation {\em hitting the planet's surface}. If the surface is subjected to a radiation-flux of $\Phi_0$, and then it changes to $\gamma \Phi_0$ (either as a result of volcanic activity or increased greenhouse gas concentration), we will simply model it by changing the solar constant's value from $S_0$ to $\gamma S_0$.

Furthermore, comparison of solutions of our system (that only differ in their initial conditions) reveals that the framework of {\em snapshot attractors}\cite{Romeiras1990, Sommerer1993} is the appropriate formulation of our problem. Practically, this means that all investigations are carried out over an {\em ensemble} of climate realizations. To our knowledge, the present study is the first in investigating the Snowball Earth transition {\em through the theory of snapshot attractors}.

The paper is organized as follows. In Section \ref{sec:setup} we introduce the climate model, and its behavior in the case of constant parameters. We also comment on the choice of the parameter drift scenario.
Then, in Section \ref{sec:individual} we present results of individual simulations obtained with differently parametrized scenarios. We demonstrate the unpredictability of individual climate realizations.
Section \ref{sec:ens} contains results of an ensemble of simulations, followed by Section \ref{sec:edge} where we discuss the climate's unstable state, the edge state. Finally, in Section \ref{sec:sum} we summarize the results and mention possible future directions of research.

\section{\label{sec:setup}The set-up}
To simulate Earth's climate system, we use the intermediate complexity climate model PlaSim \cite{Fraedrich2005a}. For the study, we use the default setup with T21 resolution, which yields a grid of approximately 5$^o$ $\times$ 5$^o$. The atmosphere is coupled to a stationary mixed layer ocean without any hydrodynamical activity, as a heat bath. The atmospheric dynamics are described by primitive equations, that represent conservation laws, thermodynamics and the hydrostatic approximation. These equations are solved on a sphere using a spectral method. With this resolution of the variables, the system has $\sim$10$^5$ degrees of freedom. Through parametrisation, the model accounts for numerous unresolved processes, including sea ice formation, which will be central in our study. For each marine cell on the grid, sea ice is allowed to form when the surface temperature is below $0~^{\circ}$C. However, ice is also assumed to be non-moving, and cannot accumulate.

\begin{figure}[h!]

    \includegraphics[width = 0.49\textwidth]{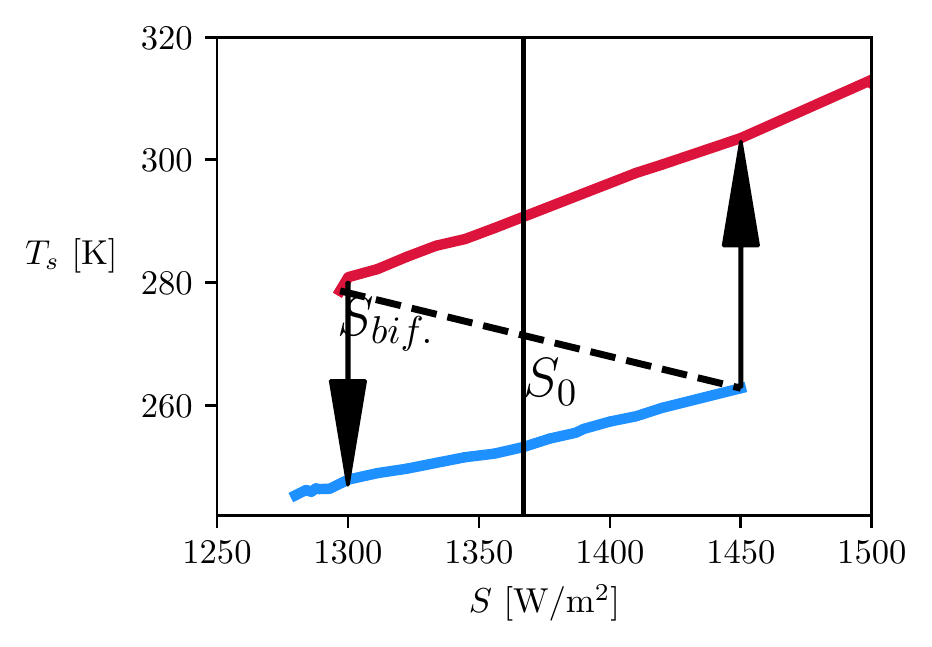} 
    \caption{The system's bifurcation diagram. In the upper panel, the northern hemisphere's average surface temperature ($T_s$) is shown, measured on July 1st, in  stable equilibrium states with the given $S$ (after transients of about 100 years are discarded). The red curve corresponds to the warm state, resembling today's climate, while blue marks the snowball state. The vertical line mark the value of $S_0$.  Dashed line indicates the unstable state (edge state). Two values, at which bifurcations occur, are indicated by arrows.}
    \label{fig:bif}
\end{figure}

Solar irradiance is treated as a boundary condition, which is mainly parametrized by the solar constant, $S$. Its value corresponding to the present climate is $S_0 = 1367$ W/m$^2$, measured at the top of the atmosphere. There were numerous studies on the relationship between the climate system and the solar constant, investigating both the thermodynamic \cite{Lucarini2010, Boschi2013} and statistical properties \cite{tantet2018}. The PlaSim model faithfully reflects the fact, that for a wide range of $S$ two alternative stable states are possible, {\em i.e. the climate is bistable}. Fig. \ref{fig:bif} shows the bifurcation diagram of our model. Note that the bifurcation diagram is constructed as usual, without any parameter drift. 

The climate's two possible stable states are attractors, these are marked with solid curves in Fig. \ref{fig:bif}. The branches of the attractors show a nearly linear relationship between the average surface temperature of the northern hemisphere ($T_s$) and the solar constant. There are two bifurcation points, at which one of the attractors lose stability. Between them, for $1297$~ W/m$^2$~$<S < 1450$ ~W/m$^2$ the system has two attractors. In a simulation, the climate's final state is determined by which basin of attraction the initial condition falls in. It is also known, that there is a third equilibrium state, which is {\em unstable} (represented by the dashed curve of Fig. \ref{fig:bif}). This is a saddle type state, which is embedded in the boundary between the two attractors basins'. It is referred to as {\em the edge state}\cite{Skufca2006, Bodai2015EBM} (or sometimes {\em Melancholia state})\cite{Lucarini2017b, lucarini2019}.


 \begin{figure}[h!]
    \centering
    \includegraphics[width = 0.5\textwidth]{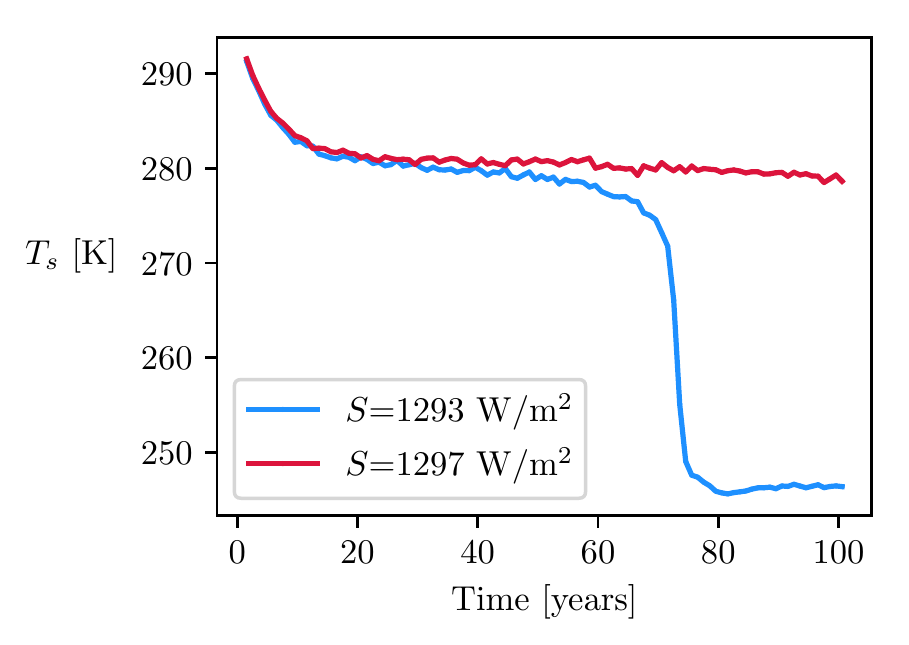}

    \caption{Two simulations which start with solar constant values near the $S_\text{bif.}$ = 1295 W/m$^2$ bifurcation point. The initial conditions are taken to be similar to today's climate. The northern hemisphere's  average surface temperature on July 1st is shown as a function of time. The simulation marked with red start with an initial condition falling into the basin of the warm attractor, while the blue one goes to the snowball state (since for that value of $S$, no warm attractor exists). The simulation time is 100 years, with year 0 signaling the start of the simulation.}
    \label{fig:conv}
\end{figure}
For our purposes, the behavior near the bifurcation point $S_\text{bif.}$ = 1295 W/m$^2$ is in the focus of interest. This is shown in Fig. \ref{fig:conv}, with the trajectories illustrating the behavior below and above the bifurcation point.  It is remarkable that the two trajectories have considerably different character. The red trajectory converges to a temperature around 280 K, with a characteristic time of 10 years. The blue trajectory also seems to follow this behavior initially, but later leaves the plateau of 280 K in a much quicker fashion, finally ending up on the snowball state's attractor, which is the only remaining stable one. Although for $S<S_{\text{bif.}}$ the warm attractor (and with it, also the edge state) has already disappeared, the initial slowing down near 280 K seems to reflect some remnants (or ghost \cite{medeiros2017}) of the saddle type unstable climate.

The bifurcation diagram presented above is a property of the system with constant parameters. Our aim is to observe a dynamical transition (tipping \cite{Lenton2008}) between the two attractors, which can be realized by introducing a parameter drift through the bifurcation point. To this end, we introduce a {\em time dependent solar constant} $S(t)$, that follows a {\em parameter drift scenario}. It is important to note that usually, tipping transitions are investigated by using slow, monotonous parameter drifts. In order to also observe the transition in the opposite
direction (that ends in the warm state), we use a qualitatively
different scenario, where the parameter is returned to its original
value \cite{Ashwin2019}. 
For the sake of simplicity, we choose a piece-wise linear function, which is described as: 
\begin{equation}
    S(t) = 	\begin{cases}
	S_0, & \text{for  } t\leq 50 \text{ year } \\
	S_0 - r (t-50 \text{ year }), & \text{for  } 50 \text{ year } < t \leq 51 \text{ year } \\
	S_0\cdot \gamma, & \text{for  } 51 \text{ year } < t \leq 101 \text{ year } \\
	S_0\cdot \gamma + r (t-101 \text{ year }), & \text{for } 101 \text{ year }< t< 102 \text{ year } \\
	S_0, & \text{for  } t\geq 102 \text{ year, } 
	\end{cases}
	\label{eq1}
\end{equation}
where $r=(1-\gamma)S_0/$ year. 
This function describes a 50 year long initial plateau of constant $S=S_0$, followed by a very quick (1 year long) linear ramp, which ends at  $S_0\gamma$ ($\gamma<1$). This value is kept constant for another 50 years, after which $S$ increases again to the value of $S_0$. The function can be seen in the lower panel of Fig. \ref{fig:gam}. 

The time dependent solar constant itself may seem ad-hoc, but it successfully incorporates multiple effects that influence solar irradiation. Furthermore, it is a standard choice of control parameter when investigating the Snowball Earth transition \cite{Budyko1969, Ghil1976, Lucarini2010, Bodai2015EBM, Lucarini2017b}. The two fast linear ramps of the $S(t)$ scenario can be interpreted as instantaneous events that trigger a Snowball transition (e.g. erupting supervolcano, meteor impact). For example, a sufficiently large volcanic eruption would decrease the surface's solar irradiation by up to an order of a few percents \cite{oppenheimer2002}. 

In the next section, we briefly present results of individual simulations. However, these turn out to be {\em not representative} of the dynamics of all the possible climate histories under the given forcing. Instead, we turn to the framework of parallel climate realizations \cite{Herein2017}, i.e. to the application of {\em snapshot attractors} \cite{Romeiras1990, Sommerer1993,Bodai2012, Drotos2015} to the climate dynamical model. 

It is clear, that in our case, a single climate realization and the ensemble yield different results. This is because the system subjected to the parameter drift thus becomes nonautonomous and is no longer ergodic. In these cases, following the ensemble of trajectories (or climate realizations) leads to a complete description of the statistics underlying the climate. In the following, we construct the changing climate's snapshot attractor and deduce the expected (or typical) behavior, along with other statistical features. 

\section{\label{sec:individual}Case studies in individual simulations}
First we investigate the impact of the solar constant's decrease, which is controlled by parameter $\gamma$. To illustrate this, we calculate the mean surface temperature of the northern hemisphere, and plot it in every year's July 1st. For simplicity, in the remainder of the paper, we will refer to this quantity simply as the {\em mean surface temperature}, $T_s$. 

\begin{figure}[h!]
    \centering
    \includegraphics[width = \linewidth]{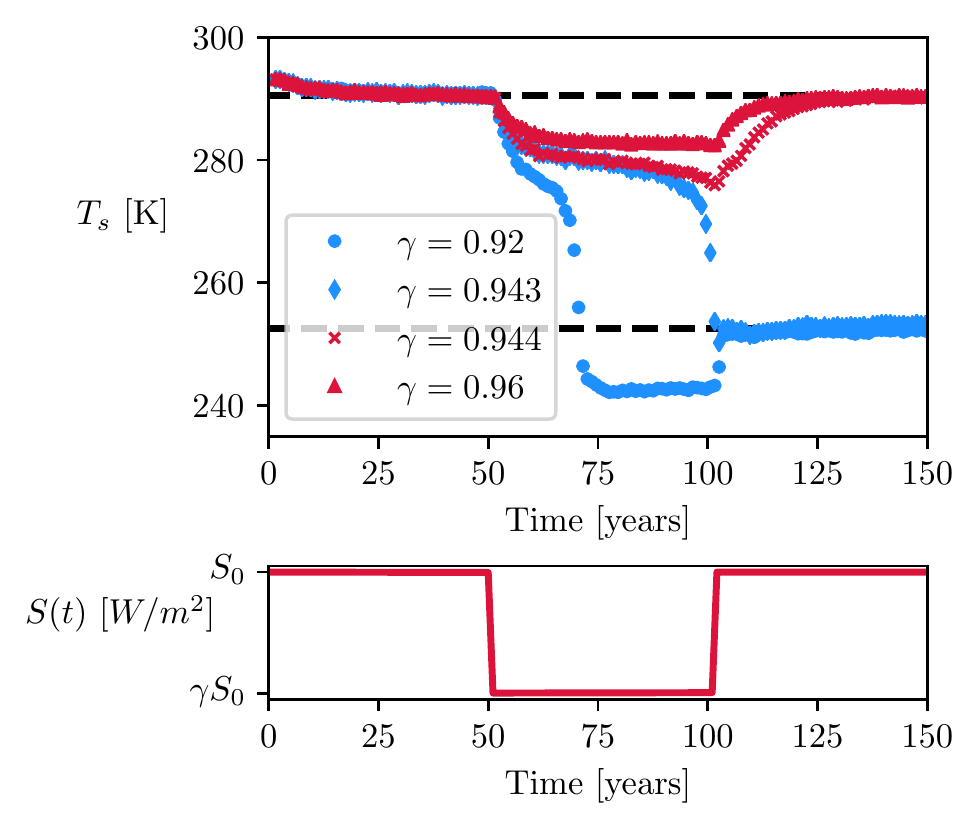}
    \caption{Upper panel: Individual simulations with different $S(t)$ parameter drift scenarios. Bottom panel: Sketch of the parameter drift scenario (\ref{eq1}). The blue curves end up in the frozen state with $\gamma = 0.92$ ($S_0\gamma = 1257.64$ W/m$^2$) and $\gamma = 0.943$ ($S_0\gamma = 1289$ W/m$^2$).
    The red curves are trajectories returning to the warm state with $\gamma = 0.944$ ($S_0\gamma = 1290.45$ W/m$^2$), $\gamma = 0.96$ ($S_0\gamma = 1312.32$ W/m$^2$). Dashed horizontal lines mark the two stable states that coexist for fixed $S=S_0$.}
    \label{fig:gam}
\end{figure}

Fig. \ref{fig:gam} shows the mean surface temperature's time dependence plotted with four values of $\gamma$. It is obvious, that if $\gamma$ is sufficiently close to 1, the solar constant $S$ is not decreased enough to make the system freeze over. This can be seen on the red curves of Fig. \ref{fig:gam}, corresponding to $\gamma = 0.96$ and $0.944$. The three plateaus of the scenario are easily discerned. $T_s$ converges, in an exponential fashion, to the equilibrium values corresponding to $S= S_0$ and $S = \gamma S_0$. 

Conversely, when $\gamma$ is much farther from 1, the mean surface temperature drops almost immediately and the system freezes over. In this case, even after the solar constant is increased back up to $S=S_0$, there is no hope to return to the warm attractor, because of the strong ice-albedo feedback.

The interesting behavior is seen for intermediate values of $\gamma$. With $\gamma = 0.943$ and $\gamma = 0.944$, the $T_s(t)$ curves are initially very similar. On the plateau corresponding to $S= \gamma S_0$ the two curves do not saturate at an equilibrium value, but their deviation from the value $T_s = 280$ K is still very slow. Then, just before the second linear ramp at $t=$100 years, the trajectories separate, with one ending up in the warm climate's attractor, and the other falling to the Snowball Earth's attractor. 

We identify this instability as the influence of the remnants of the system's unstable equilibrium, the edge state, since the phenomenon is very similar to the usual slowing down near a saddle in dynamical systems. For certain values of $\gamma$, the drifting system spends considerable amount of time near the edge state, and this causes it not to reach the equilibrium solution at $S = \gamma S_0$.  In the following, we concentrate on this threshold value of $\gamma$.

\begin{figure}[h!]
    \centering
    \includegraphics[width = \linewidth]{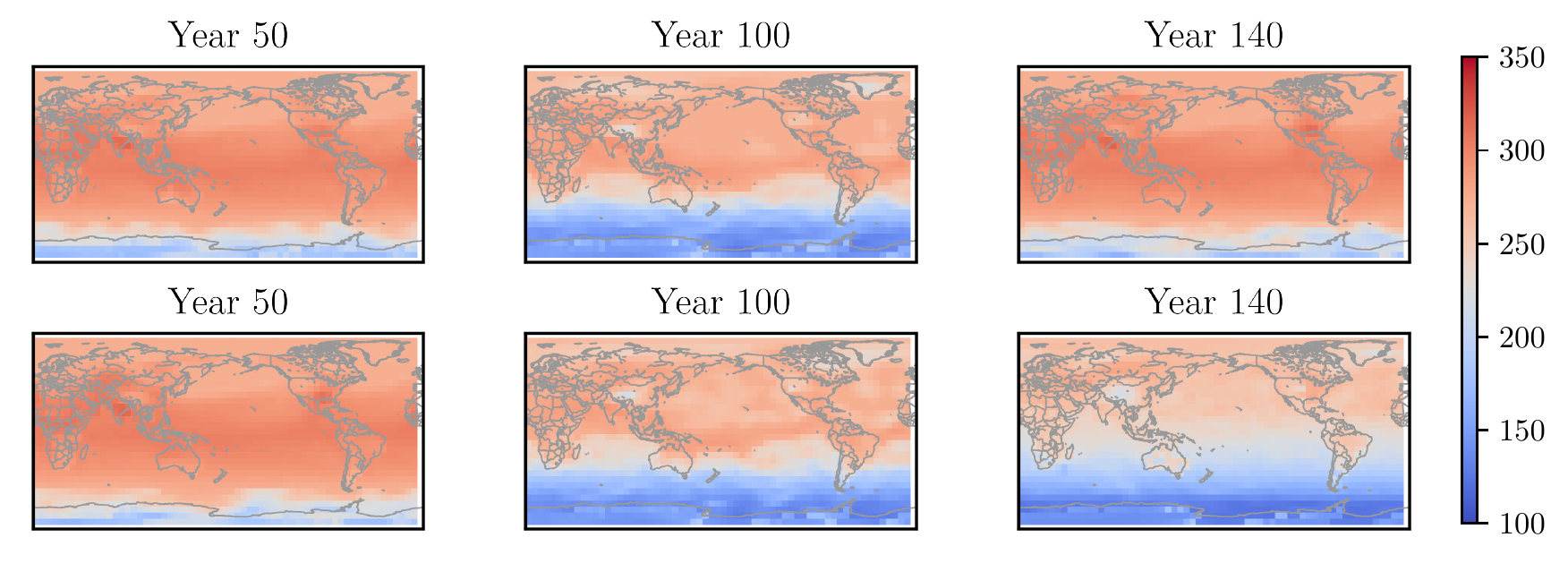}
    \caption{Surface temperature fields along two different trajectories, with $\gamma = 0.943$. We show the temperature in three time instants, in years 50, 100, and 140 after the start of the simulation on July 1st. The colors indicate temperature measured in K. In the top row, a simulation that is returning to the warm climate is shown (this is the same simulation that also appears in Fig. \ref{fig:gam}), while in the bottom row, the trajectory ends up in the Snowball state.}
   \label{fig:ts1}
  \end{figure}

Fig. \ref{fig:ts1} shows the temperature field in three different time instants, during two simulations in the same scenario (with $\gamma = 0.943$). The upper and lower panels show results of simulations started from two different initial conditions. The difference in initial conditions is realized by taking a state {\em close} to Earth's current climate (the same one as in Fig. \ref{fig:conv}), and then perturbing the pressure field in specific places. These are on the order of 10 hPa. This results in two initial states that are different, but still very close in the high dimensional phase space.

The first panel shows the temperature fields at the moment, when the solar constant starts to decrease. By this time, the system has reached its attractor (the convergence time is about 20 years, as it is also seen in Fig. \ref{fig:gam}) that exists at $S=S_0$. Notice that the two fields are very similar to each other with a warmer northern and cooler southern hemisphere due to their respective summer and winter season in July.
The middle panel shows the time instant right before the second ramp, when the solar constant starts increasing from $S=\gamma S_0$. The difference between the two fields is still barely noticeable. However, in the third panel we see some drastic differences. In the upper row, the simulation simply returns to the attractor of the warm climate. While in the bottom row, we see that at the end of the simulation, the temperatures are much lower. This shows that in the second case a {\em tipping transition} \cite{Lenton2008} to the Snowball state has occurred.

\begin{figure}[h!]
    \centering
    \includegraphics[width = \linewidth]{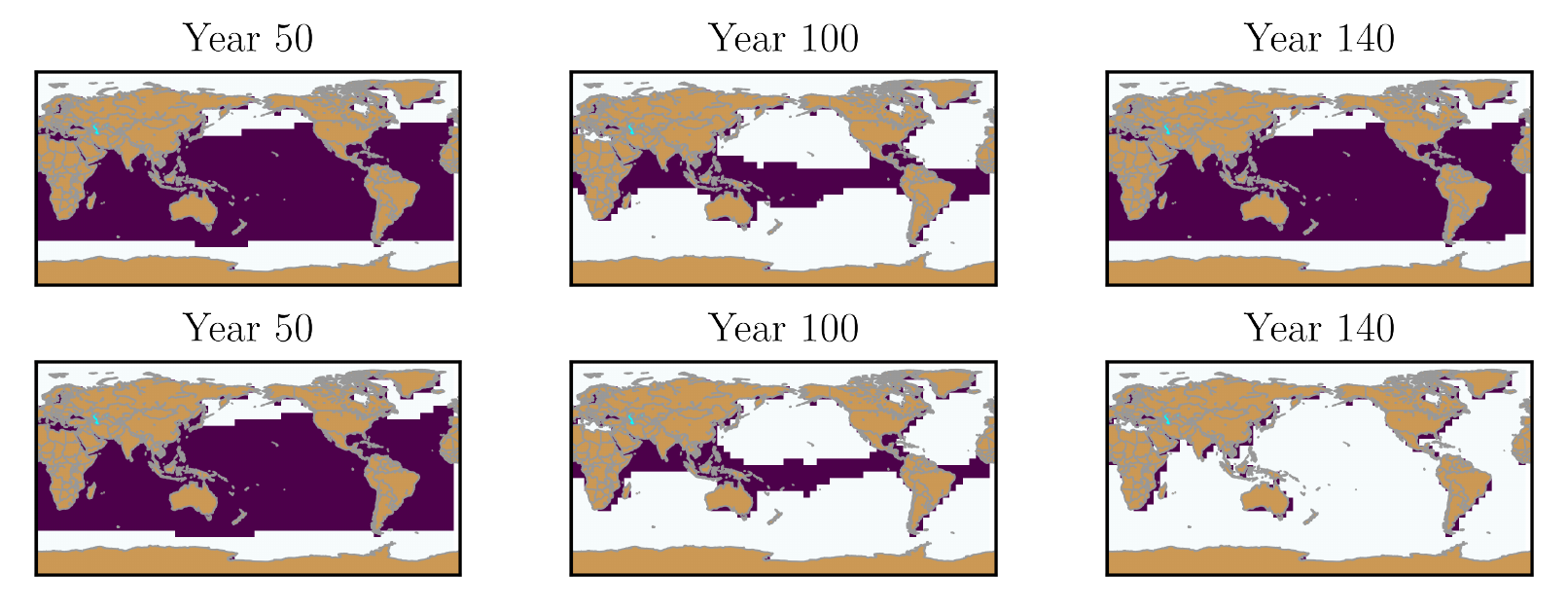}
    \caption{Sea ice cover over two different simulations. Distribution of sea ice is shown in the same time instants as Fig. \ref{fig:ts1}. The violet points indicate water at the sea surface, while white points mark the presence of surface ice. }
   \label{fig:sic1}
  \end{figure}

The same general features are perhaps more pronounced in a different representation. Figure \ref{fig:sic1} shows the distribution of sea ice in the same two simulations. By year 50, only the regions near the poles are covered in sea ice, as it is expected on the attractor belonging to $S_0$ (today's climate). Then, as the solar constant is decreased, the area of sea ice starts to grow. By year 100, the difference between the two (initially nearly identical) simulations becomes apparent: one of them has a wider band of liquid sea along the equator than the other. The tipping transition is clearly seen in terms of sea ice coverage: by year 140, in the bottom panel of Fig. \ref{fig:sic1} the simulation results in a completely frozen planet, with all of the oceans covered in ice.

\section{\label{sec:ens}The ensemble view: visualizing the snapshot attractor}
Looking at the two simulations of Fig. \ref{fig:ts1} and \ref{fig:sic1}, we see that both the warm climate and the Snowball state can be reached with the {\em same} parameter drift scenario. That is, despite the model being completely deterministic, the transition does not occur with certainty.

However, we still have no information on the {\em typical behavior} of the system. To study this question in detail, we initialize an ensemble of trajectories, and follow their time evolution. We generate different initial conditions by adding random perturbations to the pressure field, as described in the previous section. 

Following an ensemble of trajectories is becoming increasingly popular in the study of climate dynamics \cite{Pierini2016d, Herein2016b}, even in experimental studies \cite{Vincze2017}. They are referred to as {\em parallel climate realizations}, and used for extracting statistical properties. For example, one can estimate the {\em typical} behavior (the ``mean state''), and the characteristics of the fluctuations (``internal variability''). 

From a dynamical point of view, the ensemble of trajectories can be interpreted as a {\em snapshot attractor} (or {\em pullback attractor}\cite{Crauel1997, Ghil2008a, Chekroun2011a, Ghil2015} in the mathematical literature). 
After an initial convergence time, a numerical ensemble is thought to be close to the actual snapshot attractor. In this case, for any time instant, the ensemble defines a probability distribution over phase space. This is of course, also time dependent in general. For low dimensional systems, it is even possible to visualize the snapshot attractor and its natural distribution \cite{Chekroun2011a, Bodai2012, Kaszas2016}. The idea is, that all statistical measures should be determined with respect to the snapshot attractor's natural distribution. E. g. the typical, expected behavior should correspond to the distribution's average, and the internal variability can be quantified by the distribution's standard deviation. 

\begin{figure}[h!]
    \centering
    \includegraphics[width=\linewidth]{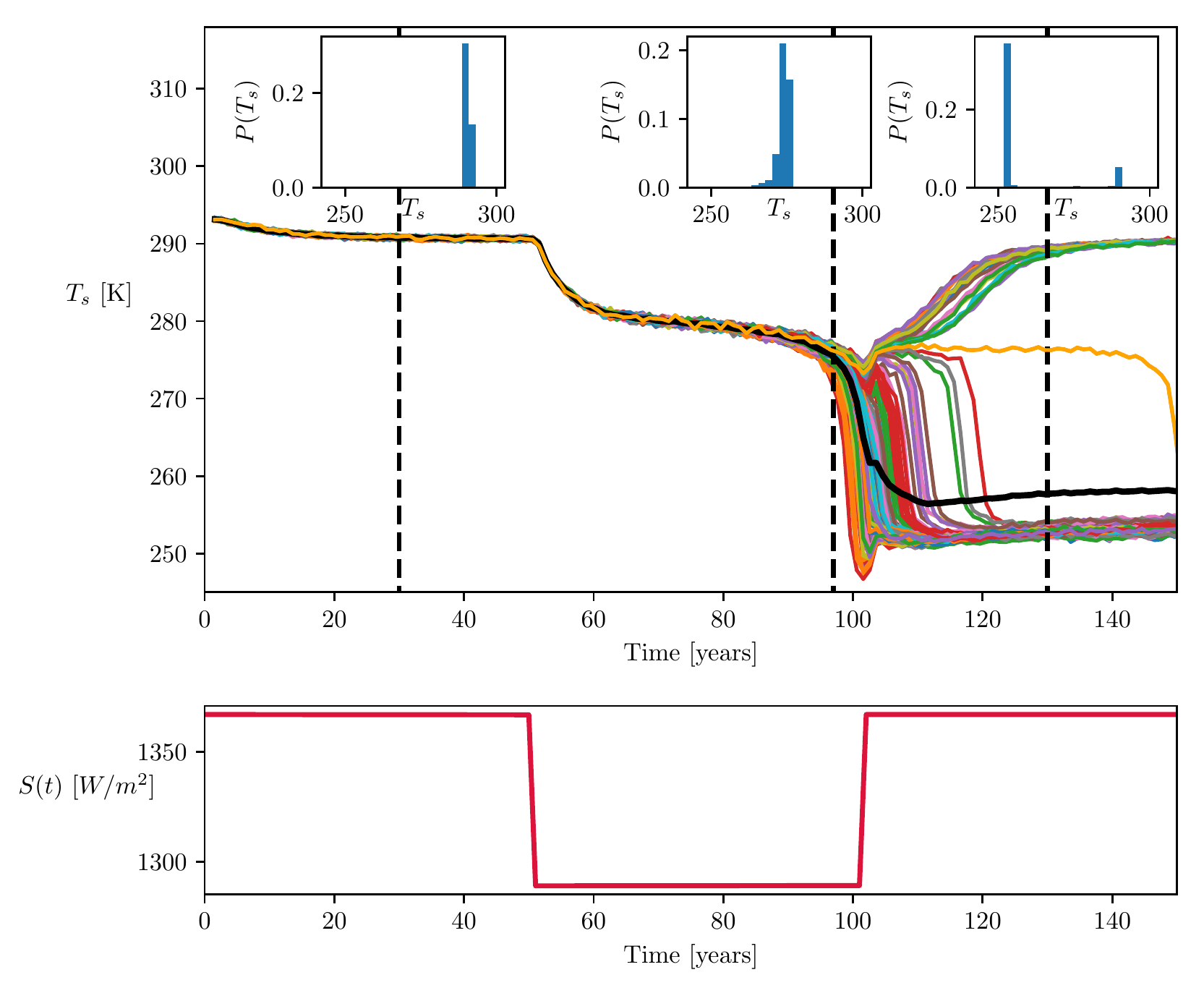}
    \caption{Top row: Time dependence of the mean surface temperature in the system that has a solar constant drifting according to Eq. (\ref{eq1}) with $\gamma = 0.943$. Each curve is a different climate realization, i.e. an ensemble of 125 members is shown, representing the drifting climate's snapshot attractor. The three insets show the probability distribution of mean surface temperature on the snapshot attractor in different time instants (marked by dashed lines). The black curve is the ensemble average of this distribution. 
    In the lower panel, the parameter drift scenario, the graph of Eq. (\ref{eq1}) is displayed.}    \label{fig:snap}
\end{figure}

Fig. \ref{fig:snap} illustrates the time evolution of the system's snapshot attractor during the parameter drift scenario. In the upper panel, the average surface temperature $T_s$ is shown in all 125 members of the ensemble. 
During the first 50 years, all trajectories converge to a unique equilibrium temperature. This temperature characterizes the warm climate with $S=S_0$. It is followed by a gradual decrease in temperature, in a similar manner as the individual simulation Fig. \ref{fig:gam} (with $\gamma = 0.943$). 

After 50 years, up until about 90 years after the start of the simulation, all of the trajectories still behave in the same way. They show a steady convergence towards the Snowball state's attractor, which is the only stable one at $S=\gamma S_0$. Here, the snapshot attractor has a very small extension in phase space, and its natural measure is a single, very narrow peak (see inset of Fig. \ref{fig:snap}).

However, when the trajectories reach the temperature values around 280 K, the peak in the distribution starts to widen. This is the temperature value, which 
is expected to be very close to the remnants of the unstable edge state, which was a part of the basin boundary between the two stable attractros, at the value $S=S_0$. The trajectories are coming to its vicinity along the remanants of the edge state's stable direction, but quickly leave along the unstable direction. It is this mechanism, that makes the snapshot attractor more extended. By the time the solar constant starts to increase again, at the 100 year mark, the natural distribution is quite wide. After the second ramp, both of the stationary attractors are stable again, and the basin boundary splits the ensemble into two parts: the trajectories start to converge to one of the two globally stable attractors at $S=S_0$. This results in either a fully ice covered Snowball Earth, or a return to the fairly warm climate. From the point of view of a single realization, it is {\em completely random}, whether it is possible to return to the state of warm climate or not. Considering the snapshot attractor reveals that the process is indeed random, but the {\em probability} of ending up on a given attractor is given by the natural distribution of the snapshot attractor. For example, if one is interested in the probability of the whole Earth freezing over, it is reasonable to talk about a {\em tipping probability} to the Snowball attractor. In our numerical study this probability is $P = 0.85$, indicating that the Snowball Earth is the more probable outcome. This is in contrast to the way tipping transitions are usually discussed in deterministic systems, where as a result of a monotonous parameter drift, either all trajectories tip, or none of them do \cite{Wieczorek2017}.

The ``typical behavior'', the ensemble average $\langle T_s \rangle$ computed over the snapshot attractor, is the black curve in Fig. \ref{fig:snap}. It is important  to note, that because the natural distribution becomes bimodal for $t>100$ years, the average lies somewhere between the two peaks (the two stable attractors). Paradoxically, the average climate is {\em almost never} realized. Instead, it is much clearer to say that because of the mechanism described above, the snapshot attractor {\em splits}, and after the second ramp, two snapshot attractors exist. Strictly speaking, the snapshot attractor remains a single object, but its two parts can be considered separately. This is because no transition is possible between them. By year 140, the two parts (practically, two snapshot attractors) reach the two attractors that are stable for the stationary value of $S=S_0$. The  averages taken with respect to the natural distributions of the two snapshot attractors describe a freezing climate, and a warm climate. 
Note, however, that the relative weights of the two disjoint parts are precisely that characterize the probability of each outcome.

\begin{figure*}[t]
    \centering
    \includegraphics[width = 0.99\textwidth]{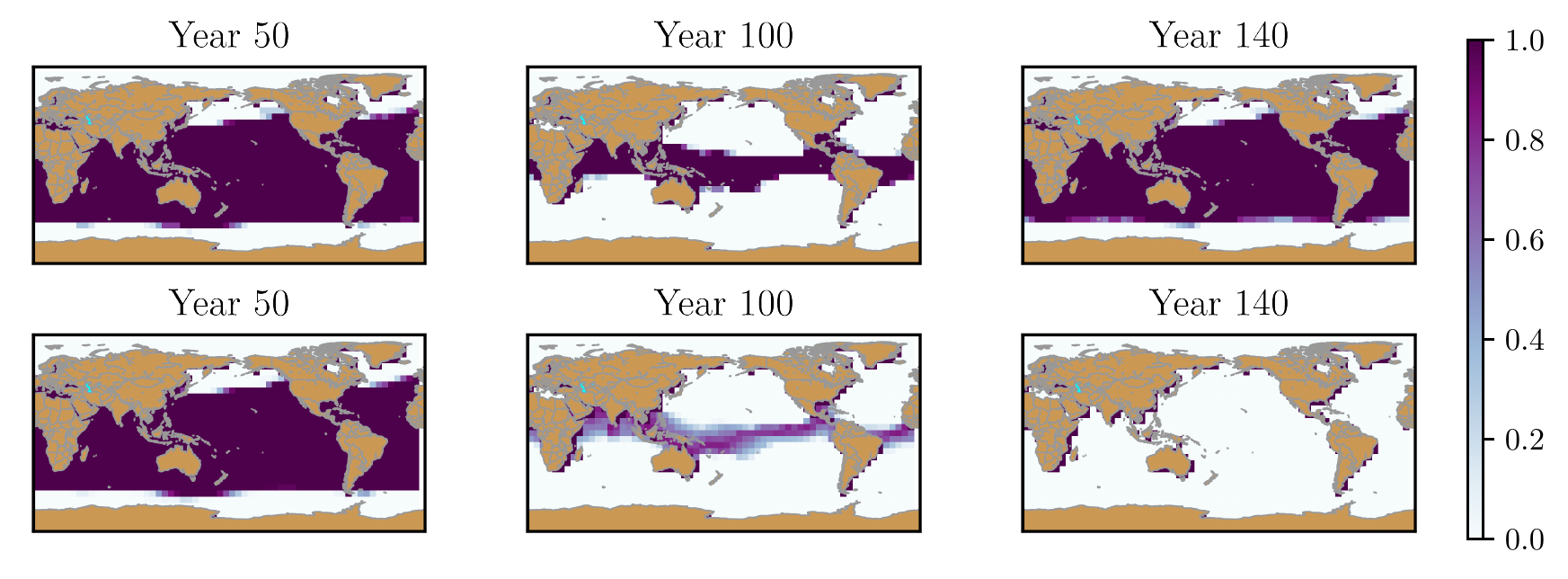}
    \caption{ Sea ice cover in the two different climates (as snapshot attractors). Distribution of sea ice is shown in the same time instants as Fig. \ref{fig:ts1}. The color coding reflects the local probability of the formation of sea ice.}
   \label{fig:sic2}
  \end{figure*}
  
  Figure \ref{fig:sic2} shows the sea ice cover in the typical freezing, and warm climates. Computing the sea ice cover in the different members of the ensemble in each location gives the probability of the sea freezing over locally. This probability is reflected in the color coding of Fig. \ref{fig:sic2}. 
The top row shows the typical behaviour, that ends in the warm state, while in the middle row, the typical snowball climate can be seen. 

Finally, it is important to notice that there is one trajectory, that has a mean surface temperature of approximately 280 K, for surprisingly long times (orange curve of Fig. \ref{fig:snap}). This means that it stays far away from the two stable attractors for a considerable time. Only after 140 years does it finally converge to the Snowball Earth's attractor. This simulation is thought to approximate the saddle-like edge state. 

\section{\label{sec:edge}The edge state between warm and Snowball climates}
The edge state has an important role in the transition to the Snowball Earth's attractor. The one (orange colored) trajectory, that stays far away from both of the attractors is thought to approximate the edge state very well.

\begin{figure}[h!]
    \centering
    \includegraphics[width = \linewidth]{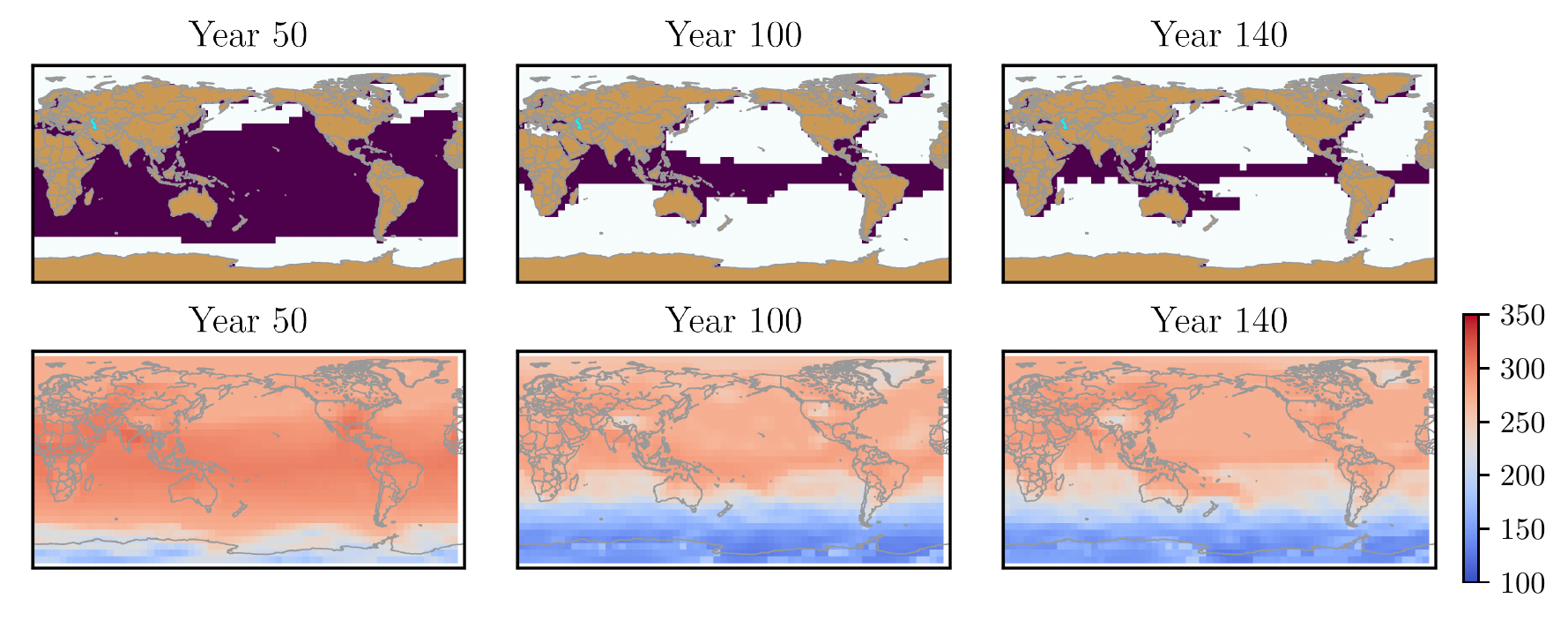}
    \caption{Climate realization that approximates the edge state. In the top row, the sea ice cover, and in the bottom row, the surface temperature are displayed. }
   \label{fig:edge}
  \end{figure}

Figure \ref{fig:edge} shows both the sea ice and surface temperature distributions of this simulation, that we will refer to as {\em the } edge state \cite{Lucarini2017b}. The surface temperature fields reflect the slight decrease of the orange curve of Fig. \ref{fig:snap}, for times between 100 and 140 years. 

It is worth comparing the sea ice distribution to that of the typical behaviors (first two rows of Fig. \ref{fig:sic2}). For example, we see that at year 100, the edge state is more similar to a climate that ends on the warm attractor, with a wide band of sea remaining at the Equator. Despite the solar constant's increase after this time, the sea ice cover is slightly increasing. In the edge state, this process stops with only a narrow band (1 or 2 grid cells wide) of water remaining at the Equator. This is the main difference between the two typical behaviors: the majority of the planet becomes ice covered, but a continuous mass of ice does not form.

For a fixed $S$, the edge state is embedded in the basin boundary between the two stable attractors. Because of the parameter drift, each member of the ensemble seems to cross the basin boundary and approach this saddle type instability. Or equivalently, we can think of the snapshot attractor itself approaching the edge state. The edge state causes the snapshot attractor to widen, because of the repulsion along the remnants of the unstable directions. When the parameter returns to the value of $S=S_0$, the members of the ensemble will fall in the basins of either of the stable attractors (or in special cases, some members fall close to the basin boundary, and stay near the edge state for long times). This mechanism leads to the probabilistic outcome of the parameter drift.

\begin{figure}[h!]
    \centering
    \includegraphics[width = 0.49\textwidth]{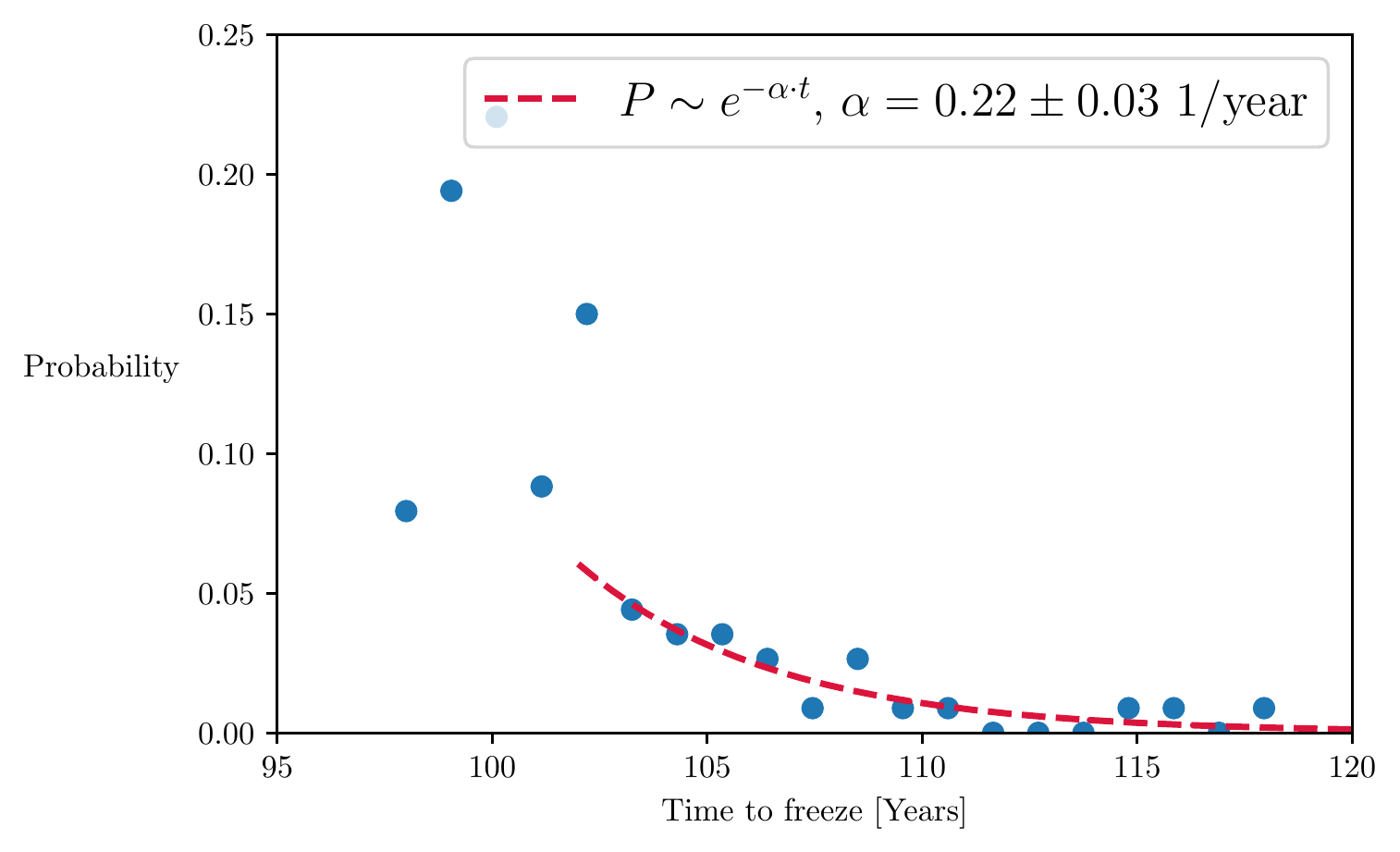}
    \caption{Distribution of freezing times within the ensemble. A realization is considered to be frozen when its average temperature falls below 260K. The probability of a simulation having a certain freezing time (with respect to the snapshot attractor's natural measure) is displayed. The dashed line shows an exponential tail fitted to the distribution, $P \sim e^{-\alpha t}$, with $\alpha = 0.22 \pm 0.03$ 1/year. }
    \label{fig:freezingtimes}
\end{figure}

A further argument in favor of the edge state's role in the Snowball Earth transition is the distribution of {\em freezing times}. We choose a value in the average surface temperature, below which a climate realization is considered to have converged to the frozen state. Then, the freezing time of a trajectory is the time instant at which it crosses this threshold temperature (i. e. the sharp drop in the curves of Fig. \ref{fig:snap}). In Fig. \ref{fig:freezingtimes} the distribution of this variable can be seen. The distribution appears to have an exponential tail, the probability decreases as $\sim e^{-\alpha t}$ for large values of the freezing time $t$. From the probability distribution of Fig. \ref{fig:freezingtimes}, we get $\alpha = 0.22$ 1/year, which implies a characteristic time scale of $1/\alpha = 4.5$ years. This exponentially decaying behaviour is characteristic of escape processes: for example the probability of staying in the neighborhood of a saddle \cite{YC2011}, or in a given region of phase space \cite{Altmann2013, Hellmann2016,Kaszas2018}.

\section{\label{sec:sum}Summary}
We investigated parameter drift induced transitions between a warm climate (resembling today's climate) and the Snowball-Earth in an intermediate complexity climate model. The system has a bistable regime in the solar constant, where the two stable attractors are separated by the unstable edge state. 

The solar constant's infinitely slow drift, corresponds to the construction of the bifurcation diagram Fig.~\ref{fig:bif}. Furthermore, we chose a continuous drift with a finite rate. A piecewise linear parameter drift scenario (defined by Eq. \ref{eq1}) was used, parametrized by the fractional decrease in solar constant, $\gamma$. Although the particular choice of scenario was a special one (in the sense that it consisted of mainly constant plateaus), the qualitative observations would remain the same for a more general choice of parameter drift. We showed that for a certain, critical value of $\gamma$, the trajectories of the system are expected to go near the edge state. From the results of Fig. \ref{fig:gam}, this value is $\gamma = 0.943$. 

Instead of investigating individual simulations, we turned to the theory of snapshot attractors, that requires following an ensemble of parallel climate realizations. The whole ensemble was subject to the parameter drift scenario with the critical $\gamma = 0.943$. The ensemble of 125 members provides an approximation to the snapshot attractor and its natural distribution. Figure \ref{fig:snap} shows a projection of the snapshot attractor, the mean surface temperature. We see that initially it has a small extension, but near the end of the plateau at $S=\gamma S_0$, the snapshot attractor grows (its distribution widens), and eventually splits. The resulting sub ensembles converge to the stationary attractors existing at $S=S_0$. We say that the two subsets of the ensemble (that converge to either attractor after 150 years) belong to two {\em different, coexisting} snapshot attractors \cite{Pierini2018}. 

The snapshot attractor's natural distribution plays an important role in prediction. In the beginning, when the system has a constant $S=S_0$, the distribution is sharp, fairly accurate predictions are possible (in the sense of the accessible range of the given variables). When the parameter starts to drift, this changes. The snapshot attractor's extension, and with it, internal variability drastically increases. At the end of the plateau of $S=\gamma S_0$, 30 K differences are possible between certain members of the ensemble. For a prescribed parameter drift scenario, this can be interpreted as being a precursor to the splitting of the snapshot attractor, which is the projection of a future catastrophic event, in the language of climate science. This is the same phenomenon that is well known in statistical physics. The phase transition (or critical transition) is usually preceded by the divergence of the ensemble's standard deviation \cite{reichl2016}.

The exact moment, when the planet freezes over is just as hard to predict. In this regard, 20 year differences are observed in the freezing times of the ensemble members. 

It is important to emphasize, that the transition from the vicinity of the Snowball state to the warm climate is possible, but only with a certain {\em probability}. This also means that on the level of single realizations, it is impossible to predict whether the system ends in a frozen, Snowball state or a warm, habitable climate. The probability of the transition depends on the snapshot attractor's natural distribution, at the time of the splitting.

Applying the theory of snapshot attractors, we were also able to obtain a climate realization that, for long times, stays near the third equilibrium climate, the edge state.
With a critical value of $\gamma$, the ensemble members (which approximate the snapshot attractor) get close to the edge state. Even with the ensemble of size 125, there was one simulation, that remained close to the unstable solution during the whole parameter drift scenario (orange curve of Fig. \ref{fig:snap}). This corresponds to a climate realization that is only partly frozen, as seen in the top row of Fig. \ref{fig:edge}. 

Many of our observations, for example the splitting of the snapshot attractor, are expected to carry over to possibly much more complex, up to date climate models.

\section{Acknowledgements}
Fruitful discussions with Tamás Bódai, Gábor Drótos, Tamás Tél and Miklós Vincze are gratefully acknowledged. We also thank Henk Dijkstra for his suggestions to the manuscript.
This paper was supported by the ÚNKP-18-4 (T. H.) and ÚNKP-18-2 (B. K.) New National Excellence Program of the Ministry of Human Capacities, by the János Bolyai Research Scholarship of the Hungarian Academy of Sciences (T. H.), and by the National Research, Development and Innovation Office – NKFIH under grants PD-121305 (T. H.), PD-124272 (M.  H.), FK-124256 and K-125171 (T. H., M. H.)

\bibliographystyle{apsrev4-1.bst}
\bibliography{references}

\begin{thebibliography}{46}%
\makeatletter
\providecommand \@ifxundefined [1]{%
 \@ifx{#1\undefined}
}%
\providecommand \@ifnum [1]{%
 \ifnum #1\expandafter \@firstoftwo
 \else \expandafter \@secondoftwo
 \fi
}%
\providecommand \@ifx [1]{%
 \ifx #1\expandafter \@firstoftwo
 \else \expandafter \@secondoftwo
 \fi
}%
\providecommand \natexlab [1]{#1}%
\providecommand \enquote  [1]{``#1''}%
\providecommand \bibnamefont  [1]{#1}%
\providecommand \bibfnamefont [1]{#1}%
\providecommand \citenamefont [1]{#1}%
\providecommand \href@noop [0]{\@secondoftwo}%
\providecommand \href [0]{\begingroup \@sanitize@url \@href}%
\providecommand \@href[1]{\@@startlink{#1}\@@href}%
\providecommand \@@href[1]{\endgroup#1\@@endlink}%
\providecommand \@sanitize@url [0]{\catcode `\\12\catcode `\$12\catcode
  `\&12\catcode `\#12\catcode `\^12\catcode `\_12\catcode `\%12\relax}%
\providecommand \@@startlink[1]{}%
\providecommand \@@endlink[0]{}%
\providecommand \url  [0]{\begingroup\@sanitize@url \@url }%
\providecommand \@url [1]{\endgroup\@href {#1}{\urlprefix }}%
\providecommand \urlprefix  [0]{URL }%
\providecommand \Eprint [0]{\href }%
\providecommand \doibase [0]{http://dx.doi.org/}%
\providecommand \selectlanguage [0]{\@gobble}%
\providecommand \bibinfo  [0]{\@secondoftwo}%
\providecommand \bibfield  [0]{\@secondoftwo}%
\providecommand \translation [1]{[#1]}%
\providecommand \BibitemOpen [0]{}%
\providecommand \bibitemStop [0]{}%
\providecommand \bibitemNoStop [0]{.\EOS\space}%
\providecommand \EOS [0]{\spacefactor3000\relax}%
\providecommand \BibitemShut  [1]{\csname bibitem#1\endcsname}%
\let\auto@bib@innerbib\@empty
\bibitem [{\citenamefont {Harland}(1964)}]{Harland1964}%
  \BibitemOpen
  \bibfield  {author} {\bibinfo {author} {\bibfnamefont {W.~B.}\ \bibnamefont
  {Harland}},\ }\href {\doibase 10.1007/BF01821169} {\bibfield  {journal}
  {\bibinfo  {journal} {Geologische Rundschau}\ }\textbf {\bibinfo {volume}
  {54}},\ \bibinfo {pages} {45} (\bibinfo {year} {1964})}\BibitemShut {NoStop}%
\bibitem [{\citenamefont {Kirschvink}(1992)}]{Kirschvink1992}%
  \BibitemOpen
  \bibfield  {author} {\bibinfo {author} {\bibfnamefont {J.}~\bibnamefont
  {Kirschvink}},\ }\href@noop {} {\emph {\bibinfo {title} {The Proterozoic
  Biosphere: A Multidisciplinary Study}}},\ edited by\ \bibinfo {editor}
  {\bibfnamefont {C.}~\bibnamefont {Schopf}, \bibfnamefont {J.~W.;~Klein}}\
  (\bibinfo  {publisher} {Cambridge University Press},\ \bibinfo {year}
  {1992})\ pp.\ \bibinfo {pages} {51--52}\BibitemShut {NoStop}%
\bibitem [{\citenamefont {Budyko}(1969)}]{Budyko1969}%
  \BibitemOpen
  \bibfield  {author} {\bibinfo {author} {\bibfnamefont {M.~I.}\ \bibnamefont
  {Budyko}},\ }\href {\doibase 10.3402/tellusa.v21i5.10109} {\bibfield
  {journal} {\bibinfo  {journal} {Tellus}\ }\textbf {\bibinfo {volume} {21}},\
  \bibinfo {pages} {611} (\bibinfo {year} {1969})}\BibitemShut {NoStop}%
\bibitem [{\citenamefont {Sellers}(1969)}]{Sellers1969}%
  \BibitemOpen
  \bibfield  {author} {\bibinfo {author} {\bibfnamefont {W.~D.}\ \bibnamefont
  {Sellers}},\ }\href {\doibase
  10.1175/1520-0450(1969)008<0392:AGCMBO>2.0.CO;2} {\bibfield  {journal}
  {\bibinfo  {journal} {Journal of Applied Meteorology}\ }\textbf {\bibinfo
  {volume} {8}},\ \bibinfo {pages} {392} (\bibinfo {year} {1969})}\BibitemShut
  {NoStop}%
\bibitem [{\citenamefont {Ghil}\ and\ \citenamefont
  {Childress}(1987)}]{Ghil1987}%
  \BibitemOpen
  \bibfield  {author} {\bibinfo {author} {\bibfnamefont {M.}~\bibnamefont
  {Ghil}}\ and\ \bibinfo {author} {\bibfnamefont {S.}~\bibnamefont
  {Childress}},\ }\href@noop {} {\emph {\bibinfo {title} {{Topics in
  Geophysical Fluid Dynamics: Atmospheric Dynamics, Dynamo Theory, and Climate
  Dynamics}}}}\ (\bibinfo  {publisher} {Springer-Verlag New York},\ \bibinfo
  {address} {New York},\ \bibinfo {year} {1987})\BibitemShut {NoStop}%
\bibitem [{\citenamefont {Pierrehumbert}\ \emph {et~al.}(2011)\citenamefont
  {Pierrehumbert}, \citenamefont {Abbot}, \citenamefont {Voigt},\ and\
  \citenamefont {Koll}}]{Pierrehumbert2011ClimateNeoproterozoic}%
  \BibitemOpen
  \bibfield  {author} {\bibinfo {author} {\bibfnamefont {R.}~\bibnamefont
  {Pierrehumbert}}, \bibinfo {author} {\bibfnamefont {D.}~\bibnamefont
  {Abbot}}, \bibinfo {author} {\bibfnamefont {A.}~\bibnamefont {Voigt}}, \ and\
  \bibinfo {author} {\bibfnamefont {D.}~\bibnamefont {Koll}},\ }\href {\doibase
  10.1146/annurev-earth-040809-152447} {\bibfield  {journal} {\bibinfo
  {journal} {Annual Review of Earth and Planetary Sciences}\ } (\bibinfo {year}
  {2011}),\ 10.1146/annurev-earth-040809-152447}\BibitemShut {NoStop}%
\bibitem [{\citenamefont {Feulner}\ \emph {et~al.}(2015)\citenamefont
  {Feulner}, \citenamefont {Hallmann},\ and\ \citenamefont
  {Kienert}}]{Feulner2015}%
  \BibitemOpen
  \bibfield  {author} {\bibinfo {author} {\bibfnamefont {G.}~\bibnamefont
  {Feulner}}, \bibinfo {author} {\bibfnamefont {C.}~\bibnamefont {Hallmann}}, \
  and\ \bibinfo {author} {\bibfnamefont {H.}~\bibnamefont {Kienert}},\ }\href
  {\doibase 10.1038/ngeo2523} {\bibfield  {journal} {\bibinfo  {journal}
  {Nature Geoscience}\ }\textbf {\bibinfo {volume} {8}},\ \bibinfo {pages}
  {659} (\bibinfo {year} {2015})}\BibitemShut {NoStop}%
\bibitem [{\citenamefont {Tziperman}\ \emph {et~al.}(2013)\citenamefont
  {Tziperman}, \citenamefont {Ashkenazy}, \citenamefont {Gildor}, \citenamefont
  {Losch}, \citenamefont {Schrag},\ and\ \citenamefont
  {Macdonald}}]{tziperman2013}%
  \BibitemOpen
  \bibfield  {author} {\bibinfo {author} {\bibfnamefont {E.}~\bibnamefont
  {Tziperman}}, \bibinfo {author} {\bibfnamefont {Y.}~\bibnamefont
  {Ashkenazy}}, \bibinfo {author} {\bibfnamefont {H.}~\bibnamefont {Gildor}},
  \bibinfo {author} {\bibfnamefont {M.}~\bibnamefont {Losch}}, \bibinfo
  {author} {\bibfnamefont {D.~P.}\ \bibnamefont {Schrag}}, \ and\ \bibinfo
  {author} {\bibfnamefont {F.~A.}\ \bibnamefont {Macdonald}},\ }\href {\doibase
  10.1038/nature11894} {\bibfield  {journal} {\bibinfo  {journal} {Nature}\
  }\textbf {\bibinfo {volume} {495}},\ \bibinfo {pages} {90} (\bibinfo {year}
  {2013})}\BibitemShut {NoStop}%
\bibitem [{\citenamefont {Robock}\ and\ \citenamefont
  {Mao}(1995)}]{Robock1995}%
  \BibitemOpen
  \bibfield  {author} {\bibinfo {author} {\bibfnamefont {A.}~\bibnamefont
  {Robock}}\ and\ \bibinfo {author} {\bibfnamefont {J.}~\bibnamefont {Mao}},\
  }\href@noop {} {\bibfield  {journal} {\bibinfo  {journal} {Journal of
  Climate}\ }\textbf {\bibinfo {volume} {8}},\ \bibinfo {pages} {1086}
  (\bibinfo {year} {1995})}\BibitemShut {NoStop}%
\bibitem [{\citenamefont {Rose}\ \emph {et~al.}(2017)\citenamefont {Rose},
  \citenamefont {Fairchild}, \citenamefont {Benn}, \citenamefont {Brocks},
  \citenamefont {Halverson}, \citenamefont {Sadler}, \citenamefont {Cohen},
  \citenamefont {Hoffman}, \citenamefont {Voigt}, \citenamefont {Donnadieu},
  \citenamefont {Ferreira}, \citenamefont {Le~Hir}, \citenamefont {Macdonald},
  \citenamefont {Creveling}, \citenamefont {Jansen}, \citenamefont {Love},
  \citenamefont {Partin}, \citenamefont {Abbot}, \citenamefont {Tziperman},
  \citenamefont {Cox}, \citenamefont {Ramstein}, \citenamefont {Erwin},
  \citenamefont {Goodman}, \citenamefont {Rose}, \citenamefont {Maloof},
  \citenamefont {Ashkenazy},\ and\ \citenamefont {Warren}}]{Rose2017}%
  \BibitemOpen
  \bibfield  {author} {\bibinfo {author} {\bibfnamefont {B.~E.~J.}\
  \bibnamefont {Rose}}, \bibinfo {author} {\bibfnamefont {I.~J.}\ \bibnamefont
  {Fairchild}}, \bibinfo {author} {\bibfnamefont {D.~I.}\ \bibnamefont {Benn}},
  \bibinfo {author} {\bibfnamefont {J.~J.}\ \bibnamefont {Brocks}}, \bibinfo
  {author} {\bibfnamefont {G.~P.}\ \bibnamefont {Halverson}}, \bibinfo {author}
  {\bibfnamefont {P.~M.}\ \bibnamefont {Sadler}}, \bibinfo {author}
  {\bibfnamefont {P.~A.}\ \bibnamefont {Cohen}}, \bibinfo {author}
  {\bibfnamefont {P.~F.}\ \bibnamefont {Hoffman}}, \bibinfo {author}
  {\bibfnamefont {A.}~\bibnamefont {Voigt}}, \bibinfo {author} {\bibfnamefont
  {Y.}~\bibnamefont {Donnadieu}}, \bibinfo {author} {\bibfnamefont
  {D.}~\bibnamefont {Ferreira}}, \bibinfo {author} {\bibfnamefont
  {G.}~\bibnamefont {Le~Hir}}, \bibinfo {author} {\bibfnamefont {F.~A.}\
  \bibnamefont {Macdonald}}, \bibinfo {author} {\bibfnamefont {J.~R.}\
  \bibnamefont {Creveling}}, \bibinfo {author} {\bibfnamefont {M.~F.}\
  \bibnamefont {Jansen}}, \bibinfo {author} {\bibfnamefont {G.~D.}\
  \bibnamefont {Love}}, \bibinfo {author} {\bibfnamefont {C.~A.}\ \bibnamefont
  {Partin}}, \bibinfo {author} {\bibfnamefont {D.~S.}\ \bibnamefont {Abbot}},
  \bibinfo {author} {\bibfnamefont {E.}~\bibnamefont {Tziperman}}, \bibinfo
  {author} {\bibfnamefont {G.~M.}\ \bibnamefont {Cox}}, \bibinfo {author}
  {\bibfnamefont {G.}~\bibnamefont {Ramstein}}, \bibinfo {author}
  {\bibfnamefont {D.~H.}\ \bibnamefont {Erwin}}, \bibinfo {author}
  {\bibfnamefont {J.~C.}\ \bibnamefont {Goodman}}, \bibinfo {author}
  {\bibfnamefont {C.~V.}\ \bibnamefont {Rose}}, \bibinfo {author}
  {\bibfnamefont {A.~C.}\ \bibnamefont {Maloof}}, \bibinfo {author}
  {\bibfnamefont {Y.}~\bibnamefont {Ashkenazy}}, \ and\ \bibinfo {author}
  {\bibfnamefont {S.~G.}\ \bibnamefont {Warren}},\ }\href {\doibase
  10.1126/sciadv.1600983} {\bibfield  {journal} {\bibinfo  {journal} {Science
  Advances}\ }\textbf {\bibinfo {volume} {3}},\ \bibinfo {pages} {e1600983}
  (\bibinfo {year} {2017})}\BibitemShut {NoStop}%
\bibitem [{\citenamefont {Russell}(2015)}]{geoengineer}%
  \BibitemOpen
  \bibfield  {author} {\bibinfo {author} {\bibfnamefont {L.~M.}\ \bibnamefont
  {Russell}},\ }\href {\doibase 10.17226/18988} {\emph {\bibinfo {title} {2015
  AAAS Annual Meeting (12-16 February 2015)}}}\ (\bibinfo {year}
  {2015})\BibitemShut {NoStop}%
\bibitem [{\citenamefont {Steinhilber}\ \emph {et~al.}(2012)\citenamefont
  {Steinhilber}, \citenamefont {Abreu}, \citenamefont {Beer}, \citenamefont
  {Brunner}, \citenamefont {Christl},\ and\ \citenamefont
  {Fischer}}]{Steinhilber2012}%
  \BibitemOpen
  \bibfield  {author} {\bibinfo {author} {\bibfnamefont {F.}~\bibnamefont
  {Steinhilber}}, \bibinfo {author} {\bibfnamefont {J.~A.}\ \bibnamefont
  {Abreu}}, \bibinfo {author} {\bibfnamefont {J.}~\bibnamefont {Beer}},
  \bibinfo {author} {\bibfnamefont {I.}~\bibnamefont {Brunner}}, \bibinfo
  {author} {\bibfnamefont {M.}~\bibnamefont {Christl}}, \ and\ \bibinfo
  {author} {\bibfnamefont {H.}~\bibnamefont {Fischer}},\ }\href {\doibase
  10.1073/pnas.1118965109} {\bibfield  {journal} {\bibinfo  {journal}
  {Proceedings of the National Academy of Sciences}\ }\textbf {\bibinfo
  {volume} {109}},\ \bibinfo {pages} {5967} (\bibinfo {year}
  {2012})}\BibitemShut {NoStop}%
\bibitem [{\citenamefont {Feulner}(2012)}]{faintsun}%
  \BibitemOpen
  \bibfield  {author} {\bibinfo {author} {\bibfnamefont {G.}~\bibnamefont
  {Feulner}},\ }\href@noop {} {\bibfield  {journal} {\bibinfo  {journal} {Rev.
  Geophys.}\ }\textbf {\bibinfo {volume} {50}} (\bibinfo {year}
  {2012})}\BibitemShut {NoStop}%
\bibitem [{\citenamefont {Romeiras}\ \emph {et~al.}(1990)\citenamefont
  {Romeiras}, \citenamefont {Grebogi},\ and\ \citenamefont
  {Ott}}]{Romeiras1990}%
  \BibitemOpen
  \bibfield  {author} {\bibinfo {author} {\bibfnamefont {F.~J.}\ \bibnamefont
  {Romeiras}}, \bibinfo {author} {\bibfnamefont {C.}~\bibnamefont {Grebogi}}, \
  and\ \bibinfo {author} {\bibfnamefont {E.}~\bibnamefont {Ott}},\ }\href
  {\doibase 10.1103/PhysRevA.41.784} {\bibfield  {journal} {\bibinfo  {journal}
  {Physical Review A}\ }\textbf {\bibinfo {volume} {41}},\ \bibinfo {pages}
  {784} (\bibinfo {year} {1990})}\BibitemShut {NoStop}%
\bibitem [{\citenamefont {Sommerer}\ and\ \citenamefont
  {Ott}(1993)}]{Sommerer1993}%
  \BibitemOpen
  \bibfield  {author} {\bibinfo {author} {\bibfnamefont {J.~C.}\ \bibnamefont
  {Sommerer}}\ and\ \bibinfo {author} {\bibfnamefont {E.}~\bibnamefont {Ott}},\
  }\href {\doibase 10.1126/science.259.5093.335} {\bibfield  {journal}
  {\bibinfo  {journal} {Science}\ }\textbf {\bibinfo {volume} {259}},\ \bibinfo
  {pages} {335} (\bibinfo {year} {1993})}\BibitemShut {NoStop}%
\bibitem [{\citenamefont {Fraedrich}\ \emph {et~al.}(2005)\citenamefont
  {Fraedrich}, \citenamefont {Jansen}, \citenamefont {Kirk}, \citenamefont
  {Luksch},\ and\ \citenamefont {Lunkeit}}]{Fraedrich2005a}%
  \BibitemOpen
  \bibfield  {author} {\bibinfo {author} {\bibfnamefont {K.}~\bibnamefont
  {Fraedrich}}, \bibinfo {author} {\bibfnamefont {H.}~\bibnamefont {Jansen}},
  \bibinfo {author} {\bibfnamefont {E.}~\bibnamefont {Kirk}}, \bibinfo {author}
  {\bibfnamefont {U.}~\bibnamefont {Luksch}}, \ and\ \bibinfo {author}
  {\bibfnamefont {F.}~\bibnamefont {Lunkeit}},\ }\href {\doibase
  10.1127/0941-2948/2005/0043} {\bibfield  {journal} {\bibinfo  {journal}
  {Meteorologische Zeitschrift}\ }\textbf {\bibinfo {volume} {14}},\ \bibinfo
  {pages} {299} (\bibinfo {year} {2005})}\BibitemShut {NoStop}%
\bibitem [{\citenamefont {Lucarini}\ \emph {et~al.}(2010)\citenamefont
  {Lucarini}, \citenamefont {Fraedrich},\ and\ \citenamefont
  {Lunkeit}}]{Lucarini2010}%
  \BibitemOpen
  \bibfield  {author} {\bibinfo {author} {\bibfnamefont {V.}~\bibnamefont
  {Lucarini}}, \bibinfo {author} {\bibfnamefont {K.}~\bibnamefont {Fraedrich}},
  \ and\ \bibinfo {author} {\bibfnamefont {F.}~\bibnamefont {Lunkeit}},\ }\href
  {\doibase 10.1002/qj.543} {\bibfield  {journal} {\bibinfo  {journal}
  {Quarterly Journal of the Royal Meteorological Society}\ }\textbf {\bibinfo
  {volume} {136}},\ \bibinfo {pages} {2} (\bibinfo {year} {2010})}\BibitemShut
  {NoStop}%
\bibitem [{\citenamefont {Boschi}\ \emph {et~al.}(2013)\citenamefont {Boschi},
  \citenamefont {Lucarini},\ and\ \citenamefont {Pascale}}]{Boschi2013}%
  \BibitemOpen
  \bibfield  {author} {\bibinfo {author} {\bibfnamefont {R.}~\bibnamefont
  {Boschi}}, \bibinfo {author} {\bibfnamefont {V.}~\bibnamefont {Lucarini}}, \
  and\ \bibinfo {author} {\bibfnamefont {S.}~\bibnamefont {Pascale}},\ }\href
  {\doibase 10.1016/j.icarus.2013.03.017} {\bibfield  {journal} {\bibinfo
  {journal} {Icarus}\ }\textbf {\bibinfo {volume} {226}},\ \bibinfo {pages}
  {1724} (\bibinfo {year} {2013})}\BibitemShut {NoStop}%
\bibitem [{\citenamefont {Tantet}\ \emph {et~al.}(2018)\citenamefont {Tantet},
  \citenamefont {Lucarini}, \citenamefont {Lunkeit},\ and\ \citenamefont
  {Dijkstra}}]{tantet2018}%
  \BibitemOpen
  \bibfield  {author} {\bibinfo {author} {\bibfnamefont {A.}~\bibnamefont
  {Tantet}}, \bibinfo {author} {\bibfnamefont {V.}~\bibnamefont {Lucarini}},
  \bibinfo {author} {\bibfnamefont {F.}~\bibnamefont {Lunkeit}}, \ and\
  \bibinfo {author} {\bibfnamefont {H.~A.}\ \bibnamefont {Dijkstra}},\ }\href
  {\doibase 10.1088/1361-6544/aaaf42} {\bibfield  {journal} {\bibinfo
  {journal} {Nonlinearity}\ }\textbf {\bibinfo {volume} {31}},\ \bibinfo
  {pages} {2221} (\bibinfo {year} {2018})}\BibitemShut {NoStop}%
\bibitem [{\citenamefont {Skufca}\ \emph {et~al.}(2006)\citenamefont {Skufca},
  \citenamefont {Yorke},\ and\ \citenamefont {Eckhardt}}]{Skufca2006}%
  \BibitemOpen
  \bibfield  {author} {\bibinfo {author} {\bibfnamefont {J.~D.}\ \bibnamefont
  {Skufca}}, \bibinfo {author} {\bibfnamefont {J.~A.}\ \bibnamefont {Yorke}}, \
  and\ \bibinfo {author} {\bibfnamefont {B.}~\bibnamefont {Eckhardt}},\ }\href
  {\doibase 10.1103/PhysRevLett.96.174101} {\bibfield  {journal} {\bibinfo
  {journal} {Physical Review Letters}\ }\textbf {\bibinfo {volume} {96}},\
  \bibinfo {pages} {5} (\bibinfo {year} {2006})}\BibitemShut {NoStop}%
\bibitem [{\citenamefont {B{\'{o}}dai}\ \emph {et~al.}(2015)\citenamefont
  {B{\'{o}}dai}, \citenamefont {Lucarini}, \citenamefont {Lunkeit},\ and\
  \citenamefont {Boschi}}]{Bodai2015EBM}%
  \BibitemOpen
  \bibfield  {author} {\bibinfo {author} {\bibfnamefont {T.}~\bibnamefont
  {B{\'{o}}dai}}, \bibinfo {author} {\bibfnamefont {V.}~\bibnamefont
  {Lucarini}}, \bibinfo {author} {\bibfnamefont {F.}~\bibnamefont {Lunkeit}}, \
  and\ \bibinfo {author} {\bibfnamefont {R.}~\bibnamefont {Boschi}},\ }\href
  {\doibase 10.1007/s00382-014-2206-5} {\bibfield  {journal} {\bibinfo
  {journal} {Climate Dynamics}\ }\textbf {\bibinfo {volume} {44}},\ \bibinfo
  {pages} {3361} (\bibinfo {year} {2015})}\BibitemShut {NoStop}%
\bibitem [{\citenamefont {Lucarini}\ and\ \citenamefont
  {B{\'{o}}dai}(2017)}]{Lucarini2017b}%
  \BibitemOpen
  \bibfield  {author} {\bibinfo {author} {\bibfnamefont {V.}~\bibnamefont
  {Lucarini}}\ and\ \bibinfo {author} {\bibfnamefont {T.}~\bibnamefont
  {B{\'{o}}dai}},\ }\href {\doibase 10.1088/1361-6544/aa6b11} {\bibfield
  {journal} {\bibinfo  {journal} {Nonlinearity}\ }\textbf {\bibinfo {volume}
  {30}},\ \bibinfo {pages} {R32} (\bibinfo {year} {2017})}\BibitemShut
  {NoStop}%
\bibitem [{\citenamefont {Lucarini}\ and\ \citenamefont
  {Bodai}(2019)}]{lucarini2019}%
  \BibitemOpen
  \bibfield  {author} {\bibinfo {author} {\bibfnamefont {V.}~\bibnamefont
  {Lucarini}}\ and\ \bibinfo {author} {\bibfnamefont {T.}~\bibnamefont
  {Bodai}},\ }\href {\doibase 10.1103/PhysRevLett.122.158701} {\bibfield
  {journal} {\bibinfo  {journal} {Physical Review Letters}\ }\textbf {\bibinfo
  {volume} {122}},\ \bibinfo {pages} {158701} (\bibinfo {year}
  {2019})}\BibitemShut {NoStop}%
\bibitem [{\citenamefont {Medeiros}\ \emph {et~al.}(2017)\citenamefont
  {Medeiros}, \citenamefont {Caldas}, \citenamefont {Baptista},\ and\
  \citenamefont {Feudel}}]{medeiros2017}%
  \BibitemOpen
  \bibfield  {author} {\bibinfo {author} {\bibfnamefont {E.~S.}\ \bibnamefont
  {Medeiros}}, \bibinfo {author} {\bibfnamefont {I.~L.}\ \bibnamefont
  {Caldas}}, \bibinfo {author} {\bibfnamefont {M.~S.}\ \bibnamefont
  {Baptista}}, \ and\ \bibinfo {author} {\bibfnamefont {U.}~\bibnamefont
  {Feudel}},\ }\href {\doibase 10.1038/srep42351} {\bibfield  {journal}
  {\bibinfo  {journal} {Scientific Reports}\ }\textbf {\bibinfo {volume} {7}}
  (\bibinfo {year} {2017}),\ 10.1038/srep42351}\BibitemShut {NoStop}%
\bibitem [{\citenamefont {Lenton}\ \emph {et~al.}(2008)\citenamefont {Lenton},
  \citenamefont {Held}, \citenamefont {Kriegler}, \citenamefont {Hall},
  \citenamefont {Lucht}, \citenamefont {Rahmstorf},\ and\ \citenamefont
  {Schellnhuber}}]{Lenton2008}%
  \BibitemOpen
  \bibfield  {author} {\bibinfo {author} {\bibfnamefont {T.~M.}\ \bibnamefont
  {Lenton}}, \bibinfo {author} {\bibfnamefont {H.}~\bibnamefont {Held}},
  \bibinfo {author} {\bibfnamefont {E.}~\bibnamefont {Kriegler}}, \bibinfo
  {author} {\bibfnamefont {J.~W.}\ \bibnamefont {Hall}}, \bibinfo {author}
  {\bibfnamefont {W.}~\bibnamefont {Lucht}}, \bibinfo {author} {\bibfnamefont
  {S.}~\bibnamefont {Rahmstorf}}, \ and\ \bibinfo {author} {\bibfnamefont
  {H.~J.}\ \bibnamefont {Schellnhuber}},\ }\href {\doibase
  10.1073/pnas.0705414105} {\bibfield  {journal} {\bibinfo  {journal}
  {Proceedings of the National Academy of Sciences}\ }\textbf {\bibinfo
  {volume} {105}},\ \bibinfo {pages} {1786} (\bibinfo {year}
  {2008})}\BibitemShut {NoStop}%
\bibitem [{\citenamefont {Alkhayuon}\ \emph {et~al.}(2019)\citenamefont
  {Alkhayuon}, \citenamefont {Ashwin}, \citenamefont {Jackson}, \citenamefont
  {Quinn},\ and\ \citenamefont {Wood}}]{Ashwin2019}%
  \BibitemOpen
  \bibfield  {author} {\bibinfo {author} {\bibfnamefont {H.}~\bibnamefont
  {Alkhayuon}}, \bibinfo {author} {\bibfnamefont {P.}~\bibnamefont {Ashwin}},
  \bibinfo {author} {\bibfnamefont {L.~C.}\ \bibnamefont {Jackson}}, \bibinfo
  {author} {\bibfnamefont {C.}~\bibnamefont {Quinn}}, \ and\ \bibinfo {author}
  {\bibfnamefont {R.~A.}\ \bibnamefont {Wood}},\ }\href
  {http://arxiv.org/abs/1901.10111} {\enquote {\bibinfo {title} {{Basin
  bifurcations, oscillatory instability and rate-induced thresholds for AMOC in
  a global oceanic box model (Preprint)}},}\ } (\bibinfo {year}
  {2019})\BibitemShut {NoStop}%
\bibitem [{\citenamefont {Ghil}(1976)}]{Ghil1976}%
  \BibitemOpen
  \bibfield  {author} {\bibinfo {author} {\bibfnamefont {M.}~\bibnamefont
  {Ghil}},\ }\href {\doibase 10.1175/1520-0469(1976)033<0003:CSFAST>2.0.CO;2}
  {\bibfield  {journal} {\bibinfo  {journal} {Journal of Atmospheric Sciences}\
  }\textbf {\bibinfo {volume} {33}},\ \bibinfo {pages} {3} (\bibinfo {year}
  {1976})}\BibitemShut {NoStop}%
\bibitem [{\citenamefont {Oppenheimer}(2002)}]{oppenheimer2002}%
  \BibitemOpen
  \bibfield  {author} {\bibinfo {author} {\bibfnamefont {C.}~\bibnamefont
  {Oppenheimer}},\ }\href {\doibase 10.1016/S0277-3791(01)00154-8} {\bibfield
  {journal} {\bibinfo  {journal} {Quaternary Science Reviews}\ }\textbf
  {\bibinfo {volume} {21}},\ \bibinfo {pages} {1593} (\bibinfo {year}
  {2002})}\BibitemShut {NoStop}%
\bibitem [{\citenamefont {Herein}\ \emph {et~al.}(2017)\citenamefont {Herein},
  \citenamefont {Dr{\'{o}}tos}, \citenamefont {Haszpra}, \citenamefont
  {M{\'{a}}rfy},\ and\ \citenamefont {T{\'{e}}l}}]{Herein2017}%
  \BibitemOpen
  \bibfield  {author} {\bibinfo {author} {\bibfnamefont {M.}~\bibnamefont
  {Herein}}, \bibinfo {author} {\bibfnamefont {G.}~\bibnamefont
  {Dr{\'{o}}tos}}, \bibinfo {author} {\bibfnamefont {T.}~\bibnamefont
  {Haszpra}}, \bibinfo {author} {\bibfnamefont {J.}~\bibnamefont
  {M{\'{a}}rfy}}, \ and\ \bibinfo {author} {\bibfnamefont {T.}~\bibnamefont
  {T{\'{e}}l}},\ }\href {\doibase 10.1038/srep44529} {\bibfield  {journal}
  {\bibinfo  {journal} {Scientific Reports}\ }\textbf {\bibinfo {volume} {7}},\
  \bibinfo {pages} {44529} (\bibinfo {year} {2017})}\BibitemShut {NoStop}%
\bibitem [{\citenamefont {B{\'{o}}dai}\ and\ \citenamefont
  {T{\'{e}}l}(2012)}]{Bodai2012}%
  \BibitemOpen
  \bibfield  {author} {\bibinfo {author} {\bibfnamefont {T.}~\bibnamefont
  {B{\'{o}}dai}}\ and\ \bibinfo {author} {\bibfnamefont {T.}~\bibnamefont
  {T{\'{e}}l}},\ }\href {\doibase 10.1063/1.3697984} {\bibfield  {journal}
  {\bibinfo  {journal} {Chaos}\ }\textbf {\bibinfo {volume} {22}},\ \bibinfo
  {pages} {023110} (\bibinfo {year} {2012})}\BibitemShut {NoStop}%
\bibitem [{\citenamefont {Dr{\'{o}}tos}\ \emph {et~al.}(2015)\citenamefont
  {Dr{\'{o}}tos}, \citenamefont {B{\'{o}}dai},\ and\ \citenamefont
  {T{\'{e}}l}}]{Drotos2015}%
  \BibitemOpen
  \bibfield  {author} {\bibinfo {author} {\bibfnamefont {G.}~\bibnamefont
  {Dr{\'{o}}tos}}, \bibinfo {author} {\bibfnamefont {T.}~\bibnamefont
  {B{\'{o}}dai}}, \ and\ \bibinfo {author} {\bibfnamefont {T.}~\bibnamefont
  {T{\'{e}}l}},\ }\href {\doibase 10.1175/JCLI-D-14-00459.1} {\bibfield
  {journal} {\bibinfo  {journal} {Journal of Climate}\ }\textbf {\bibinfo
  {volume} {28}},\ \bibinfo {pages} {3275} (\bibinfo {year}
  {2015})}\BibitemShut {NoStop}%
\bibitem [{\citenamefont {Pierini}\ \emph {et~al.}(2016)\citenamefont
  {Pierini}, \citenamefont {Ghil},\ and\ \citenamefont
  {Chekroun}}]{Pierini2016d}%
  \BibitemOpen
  \bibfield  {author} {\bibinfo {author} {\bibfnamefont {S.}~\bibnamefont
  {Pierini}}, \bibinfo {author} {\bibfnamefont {M.}~\bibnamefont {Ghil}}, \
  and\ \bibinfo {author} {\bibfnamefont {M.~D.}\ \bibnamefont {Chekroun}},\
  }\href {\doibase 10.1175/JCLI-D-15-0848.1} {\bibfield  {journal} {\bibinfo
  {journal} {Journal of Climate}\ }\textbf {\bibinfo {volume} {29}},\ \bibinfo
  {pages} {4185} (\bibinfo {year} {2016})}\BibitemShut {NoStop}%
\bibitem [{\citenamefont {Herein}\ \emph {et~al.}(2016)\citenamefont {Herein},
  \citenamefont {M{\'{a}}rfy}, \citenamefont {Dr{\'{o}}tos},\ and\
  \citenamefont {T{\'{e}}l}}]{Herein2016b}%
  \BibitemOpen
  \bibfield  {author} {\bibinfo {author} {\bibfnamefont {M.}~\bibnamefont
  {Herein}}, \bibinfo {author} {\bibfnamefont {J.}~\bibnamefont {M{\'{a}}rfy}},
  \bibinfo {author} {\bibfnamefont {G.}~\bibnamefont {Dr{\'{o}}tos}}, \ and\
  \bibinfo {author} {\bibfnamefont {T.}~\bibnamefont {T{\'{e}}l}},\ }\href
  {\doibase 10.1175/JCLI-D-15-0353.1} {\bibfield  {journal} {\bibinfo
  {journal} {Journal of Climate}\ }\textbf {\bibinfo {volume} {29}},\ \bibinfo
  {pages} {259} (\bibinfo {year} {2016})}\BibitemShut {NoStop}%
\bibitem [{\citenamefont {Vincze}\ \emph {et~al.}(2017)\citenamefont {Vincze},
  \citenamefont {Borcia},\ and\ \citenamefont {Harlander}}]{Vincze2017}%
  \BibitemOpen
  \bibfield  {author} {\bibinfo {author} {\bibfnamefont {M.}~\bibnamefont
  {Vincze}}, \bibinfo {author} {\bibfnamefont {I.~D.}\ \bibnamefont {Borcia}},
  \ and\ \bibinfo {author} {\bibfnamefont {U.}~\bibnamefont {Harlander}},\
  }\href {\doibase 10.1038/s41598-017-00319-0} {\bibfield  {journal} {\bibinfo
  {journal} {Scientific Reports}\ }\textbf {\bibinfo {volume} {7}},\ \bibinfo
  {pages} {254} (\bibinfo {year} {2017})}\BibitemShut {NoStop}%
\bibitem [{\citenamefont {Crauel}\ \emph {et~al.}(1997)\citenamefont {Crauel},
  \citenamefont {Debussche},\ and\ \citenamefont {Flandoli}}]{Crauel1997}%
  \BibitemOpen
  \bibfield  {author} {\bibinfo {author} {\bibfnamefont {H.}~\bibnamefont
  {Crauel}}, \bibinfo {author} {\bibfnamefont {A.}~\bibnamefont {Debussche}}, \
  and\ \bibinfo {author} {\bibfnamefont {F.}~\bibnamefont {Flandoli}},\ }\href
  {\doibase 10.1007/BF02219225} {\bibfield  {journal} {\bibinfo  {journal}
  {Journal of Dynamics and Differential Equations}\ }\textbf {\bibinfo {volume}
  {9}},\ \bibinfo {pages} {307} (\bibinfo {year} {1997})}\BibitemShut {NoStop}%
\bibitem [{\citenamefont {Ghil}\ \emph {et~al.}(2008)\citenamefont {Ghil},
  \citenamefont {Chekroun},\ and\ \citenamefont {Simonnet}}]{Ghil2008a}%
  \BibitemOpen
  \bibfield  {author} {\bibinfo {author} {\bibfnamefont {M.}~\bibnamefont
  {Ghil}}, \bibinfo {author} {\bibfnamefont {M.~D.}\ \bibnamefont {Chekroun}},
  \ and\ \bibinfo {author} {\bibfnamefont {E.}~\bibnamefont {Simonnet}},\
  }\href {\doibase 10.1016/j.physd.2008.03.036} {\bibfield  {journal} {\bibinfo
   {journal} {Physica D: Nonlinear Phenomena}\ }\textbf {\bibinfo {volume}
  {237}},\ \bibinfo {pages} {2111} (\bibinfo {year} {2008})}\BibitemShut
  {NoStop}%
\bibitem [{\citenamefont {Chekroun}\ \emph {et~al.}(2011)\citenamefont
  {Chekroun}, \citenamefont {Simonnet},\ and\ \citenamefont
  {Ghil}}]{Chekroun2011a}%
  \BibitemOpen
  \bibfield  {author} {\bibinfo {author} {\bibfnamefont {M.~D.}\ \bibnamefont
  {Chekroun}}, \bibinfo {author} {\bibfnamefont {E.}~\bibnamefont {Simonnet}},
  \ and\ \bibinfo {author} {\bibfnamefont {M.}~\bibnamefont {Ghil}},\ }\href
  {\doibase 10.1016/j.physd.2011.06.005} {\bibfield  {journal} {\bibinfo
  {journal} {Physica D: Nonlinear Phenomena}\ }\textbf {\bibinfo {volume}
  {240}},\ \bibinfo {pages} {1685} (\bibinfo {year} {2011})}\BibitemShut
  {NoStop}%
\bibitem [{\citenamefont {Ghil}(2016)}]{Ghil2015}%
  \BibitemOpen
  \bibfield  {author} {\bibinfo {author} {\bibfnamefont {M.}~\bibnamefont
  {Ghil}},\ }in\ \href {\doibase 10.1142/9789814579933{\_}0002} {\emph
  {\bibinfo {booktitle} {Climate Change: Multidecadal and Beyond}}}\ (\bibinfo
  {publisher} {World Scientific},\ \bibinfo {year} {2016})\ pp.\ \bibinfo
  {pages} {31--51}\BibitemShut {NoStop}%
\bibitem [{\citenamefont {Kasz{\'{a}}s}\ \emph {et~al.}(2016)\citenamefont
  {Kasz{\'{a}}s}, \citenamefont {Feudel},\ and\ \citenamefont
  {T{\'{e}}l}}]{Kaszas2016}%
  \BibitemOpen
  \bibfield  {author} {\bibinfo {author} {\bibfnamefont {B.}~\bibnamefont
  {Kasz{\'{a}}s}}, \bibinfo {author} {\bibfnamefont {U.}~\bibnamefont
  {Feudel}}, \ and\ \bibinfo {author} {\bibfnamefont {T.}~\bibnamefont
  {T{\'{e}}l}},\ }\href {\doibase 10.1103/PhysRevE.94.062221} {\bibfield
  {journal} {\bibinfo  {journal} {Physical Review E}\ }\textbf {\bibinfo
  {volume} {94}},\ \bibinfo {pages} {062221} (\bibinfo {year}
  {2016})}\BibitemShut {NoStop}%
\bibitem [{\citenamefont {Ashwin}\ \emph {et~al.}(2017)\citenamefont {Ashwin},
  \citenamefont {Perryman},\ and\ \citenamefont {Wieczorek}}]{Wieczorek2017}%
  \BibitemOpen
  \bibfield  {author} {\bibinfo {author} {\bibfnamefont {P.}~\bibnamefont
  {Ashwin}}, \bibinfo {author} {\bibfnamefont {C.}~\bibnamefont {Perryman}}, \
  and\ \bibinfo {author} {\bibfnamefont {S.}~\bibnamefont {Wieczorek}},\ }\href
  {\doibase 10.1088/1361-6544/aa675b} {\bibfield  {journal} {\bibinfo
  {journal} {Nonlinearity}\ }\textbf {\bibinfo {volume} {30}},\ \bibinfo
  {pages} {2185} (\bibinfo {year} {2017})}\BibitemShut {NoStop}%
\bibitem [{\citenamefont {Lai}\ and\ \citenamefont {T{\'{e}}l}(2011)}]{YC2011}%
  \BibitemOpen
  \bibfield  {author} {\bibinfo {author} {\bibfnamefont {Y.-C.}\ \bibnamefont
  {Lai}}\ and\ \bibinfo {author} {\bibfnamefont {T.}~\bibnamefont
  {T{\'{e}}l}},\ }\href {\doibase 10.1007/978-1-4419-7055-8} {\emph {\bibinfo
  {title} {{Transient Chaos}}}},\ Vol.\ \bibinfo {volume} {173}\ (\bibinfo
  {publisher} {Springer},\ \bibinfo {address} {New York},\ \bibinfo {year}
  {2011})\BibitemShut {NoStop}%
\bibitem [{\citenamefont {Altmann}\ \emph {et~al.}(2013)\citenamefont
  {Altmann}, \citenamefont {Portela},\ and\ \citenamefont
  {T{\'{e}}l}}]{Altmann2013}%
  \BibitemOpen
  \bibfield  {author} {\bibinfo {author} {\bibfnamefont {E.~G.}\ \bibnamefont
  {Altmann}}, \bibinfo {author} {\bibfnamefont {J.~S.~E.}\ \bibnamefont
  {Portela}}, \ and\ \bibinfo {author} {\bibfnamefont {T.}~\bibnamefont
  {T{\'{e}}l}},\ }\href {\doibase 10.1103/RevModPhys.85.869} {\bibfield
  {journal} {\bibinfo  {journal} {Reviews of Modern Physics}\ }\textbf
  {\bibinfo {volume} {85}},\ \bibinfo {pages} {869} (\bibinfo {year}
  {2013})}\BibitemShut {NoStop}%
\bibitem [{\citenamefont {Hellmann}\ \emph {et~al.}(2016)\citenamefont
  {Hellmann}, \citenamefont {Schultz}, \citenamefont {Grabow}, \citenamefont
  {Heitzig},\ and\ \citenamefont {Kurths}}]{Hellmann2016}%
  \BibitemOpen
  \bibfield  {author} {\bibinfo {author} {\bibfnamefont {F.}~\bibnamefont
  {Hellmann}}, \bibinfo {author} {\bibfnamefont {P.}~\bibnamefont {Schultz}},
  \bibinfo {author} {\bibfnamefont {C.}~\bibnamefont {Grabow}}, \bibinfo
  {author} {\bibfnamefont {J.}~\bibnamefont {Heitzig}}, \ and\ \bibinfo
  {author} {\bibfnamefont {J.}~\bibnamefont {Kurths}},\ }\href {\doibase
  10.1038/srep29654} {\bibfield  {journal} {\bibinfo  {journal} {Scientific
  Reports}\ }\textbf {\bibinfo {volume} {6}},\ \bibinfo {pages} {29654}
  (\bibinfo {year} {2016})}\BibitemShut {NoStop}%
\bibitem [{\citenamefont {Kasz{\'{a}}s}\ \emph {et~al.}(2018)\citenamefont
  {Kasz{\'{a}}s}, \citenamefont {Feudel},\ and\ \citenamefont
  {T{\'{e}}l}}]{Kaszas2018}%
  \BibitemOpen
  \bibfield  {author} {\bibinfo {author} {\bibfnamefont {B.}~\bibnamefont
  {Kasz{\'{a}}s}}, \bibinfo {author} {\bibfnamefont {U.}~\bibnamefont
  {Feudel}}, \ and\ \bibinfo {author} {\bibfnamefont {T.}~\bibnamefont
  {T{\'{e}}l}},\ }\href {\doibase 10.1063/1.5013336} {\bibfield  {journal}
  {\bibinfo  {journal} {Chaos}\ }\textbf {\bibinfo {volume} {28}},\ \bibinfo
  {pages} {033612} (\bibinfo {year} {2018})}\BibitemShut {NoStop}%
\bibitem [{\citenamefont {Pierini}\ \emph {et~al.}(2018)\citenamefont
  {Pierini}, \citenamefont {Chekroun},\ and\ \citenamefont
  {Ghil}}]{Pierini2018}%
  \BibitemOpen
  \bibfield  {author} {\bibinfo {author} {\bibfnamefont {S.}~\bibnamefont
  {Pierini}}, \bibinfo {author} {\bibfnamefont {M.~D.}\ \bibnamefont
  {Chekroun}}, \ and\ \bibinfo {author} {\bibfnamefont {M.}~\bibnamefont
  {Ghil}},\ }\href {\doibase 10.5194/npg-25-671-2018} {\bibfield  {journal}
  {\bibinfo  {journal} {Nonlinear Processes in Geophysics}\ }\textbf {\bibinfo
  {volume} {25}},\ \bibinfo {pages} {671} (\bibinfo {year} {2018})}\BibitemShut
  {NoStop}%
\bibitem [{\citenamefont {Reichl}(2016)}]{reichl2016}%
  \BibitemOpen
  \bibfield  {author} {\bibinfo {author} {\bibfnamefont {E.}~\bibnamefont
  {Reichl}, \bibfnamefont {Linda}},\ }\href@noop {} {\emph {\bibinfo {title}
  {{A Modern Course in Statistical Physics}}}}\ (\bibinfo  {publisher}
  {Wiley-WCH},\ \bibinfo {year} {2016})\BibitemShut {NoStop}%
\end{thebibliography}%

\end{document}